\documentclass[prb,twocolumn,showpacs]{revtex4}
\usepackage{graphicx}
\usepackage{dcolumn}
\usepackage{bm}
\usepackage{amsmath}

\begin{document}

\title{Multi-reference symmetry-projected variational approaches for
  ground and excited states of the one-dimensional Hubbard model}

\author{R. Rodr\'{\i}guez-Guzm\'an$^{1,2}$, Carlos
  A. Jim\'enez-Hoyos$^{1}$, R. Schutski$^{1}$ and Gustavo
  E. Scuseria$^{1,2}$}
\affiliation{$^{1}$ Department of Chemistry, Rice University, Houston,
  Texas 77005, USA \\ $^{2}$ Department of Physics and Astronomy, Rice
  University, Houston, Texas 77005, USA }
\date{\today}

\begin{abstract}
We present a multi-reference configuration mixing scheme for
describing ground and excited states, with well defined spin and space
group symmetry quantum numbers, of the one-dimensional Hubbard model
with nearest-neighbor hopping and periodic boundary conditions. Within
this scheme, each state is expanded in terms of non-orthogonal and
variationally determined symmetry-projected configurations. The
results for lattices up to 30 and 50 sites compare well with the exact
Lieb-Wu solutions as well as with results from other state-of-the-art
approximations. In addition to spin-spin correlation functions in real
space and magnetic structure factors, we present results for spectral
functions and density of states computed with an ansatz whose quality
can be well-controlled by the number of symmetry-projected
configurations used to approximate the systems with $N_{e}$ and
$N_{e} \pm 1$ electrons. The intrinsic symmetry-broken determinants
resulting from the variational calculations have rich structures in
terms of defects 
that can be
regarded as basic units of quantum fluctuations. Given the quality of
the results here reported, as well as the parallelization properties
of the considered scheme, we believe that symmetry-projection
techniques, which have found ample applications in nuclear structure
physics, deserve further attention in the study of low-dimensional
correlated many-electron systems.
\end{abstract}

\maketitle

\section{Introduction.}

Studies of correlations arising from electron-electron interactions
remain a central theme in condensed matter physis
\cite{Dagotto-review} to better understand challenging phenomena such
as high-Tc superconductivity \cite{HTCSC-1} or colossal magnetic
resistance. \cite{Science-Dagotto} There is a need for better
theoretical models that can account for relevant correlations in
ground and excites states of fermionic systems with as much simplicity
as possible. Within this context, the repulsive Hubbard Hamiltonian
\cite{Hubbard-model_def1} has received a lot of attention since it is
considered the generic model of strongly correlated electron systems.
\cite{Dagotto-review} Hubbard-like models have also received renewed
attention in the study of cold fermionic atoms in optical lattices
\cite{optical-1} and  the electronic properties of
graphene. \cite{CastroNeto-review} 

Unlike the one-dimensional (1D) Hubbard model, which is exactly
solvable \cite{LIEB} using the Bethe ansatz, \cite{BETHE} an exact
solution of the two-dimensional (2D) problem is not known. Therefore,
it is highly desirable to develop approximations that, on the one
hand, can capture the main features of the exact 1D Bethe solution
and, on the other hand, can be extended to higher dimensions. For
small lattices, one can resort to exact diagonalization using the
Lanczos method. \cite{Dagotto-review,Lanczos-Fano} For larger systems,
several other methods have been extensively used to study the 1D and
2D Hubbard models as well as their strong coupling
versions. \cite{text-Hubbard-1D} Among such approximations, we have
the quantum Monte Carlo, \cite{Nightingale,Raedt-MC} the variational
Monte Carlo, \cite{Neuscamman-2012} the density matrix renormalization
group \cite{DMRG-White,Dukelsky-Pittel-RPP,Scholl-RMP,Xiang} as well
as approximations based on matrix product and
tensor network  states. \cite{TNPS-1} Both the dynamical mean
field theory and its cluster extensions
\cite{Zgid,DMFT-1,maier2005,stanescu2006,moukouri2001,huscroft2001,aryanpour2003,DVP-2}
have made important contributions to our present knowledge of the
Hubbard model. Other embedding approaches are also
available. \cite{Knizia-Chan} Finally, we refer the reader to the
recent state-of-the-art applications of the coupled cluster method to
frustrated Hubbard-like models. \cite{Bishop-1,Bishop-2}

Although routinely used in nuclear structure physics, especially
within the Generator Coordinate Method,
\cite{rayner-GCM-paper,rayner-GCM-parity} symmetry restoration via
projection techniques \cite{rs} has received little attention in
condensed matter physics. Nevertheless, these techniques offer an
alternative for obtaining accurate correlated wave functions that
respect the symmetries of the considered many-fermion problem. The key
idea is,\cite{rs} on the one hand, to consider a mean-field trial
state $| {\mathcal{D}} \rangle$ which deliberately breaks several
symmetries of the original Hamiltonian. On the other hand, the
Goldstone manifold $\hat{R} | {\mathcal{D}} \rangle$, where $%
\hat{R}$ represents a symmetry operation, is degenerate and the
superposition of such Goldstone states, \cite{rs} can be used to
recover the desired symmetry by means of a self-consistent
variation-after-projection procedure. \cite{rs,Rayner-Carlo-CM-1} Such
a single-reference (SR) scheme provides the optimal
Ritz-variational\cite{Blaizot-Ripka} representation of a given state
by means of only one symmetry-projected mean-field configuration. This
kind of SR variation-after-projection scheme, has already been applied
to the 1D and 2D Hubbard models
\cite{Carlos-Hubbard-1D,Rayner-2D-Hubbard-PRB-2012} as well as in
quantum chemistry within the framework of the Projected Quasiparticle
Theory. \cite{PQT-reference-1,PQT-reference-2,PQT-reference-3}

One of the main advantages of the symmetry-projected approximations
\cite{rs,Carlos-Hubbard-1D,Rayner-2D-Hubbard-PRB-2012,PQT-reference-1,PQT-reference-2,PQT-reference-3}
is that they offer compact wave functions as well as a systematic way
to improve their quality by adopting a multi-reference (MR)
approach. In this case, a set of symmetry-broken mean-field states $|
{\mathcal{D}}^{i} \rangle $ is used to build Goldstone manifolds
$\hat{R} | {\mathcal{D}}^{i} \rangle$ whose superposition can be used
to recover the desired symmetries of the
Hamiltonian. \cite{Carlo-review,Fukutome-original-RSHF} The key idea
is then to expand a given state in terms of several symmetry-projected
and variationally determined mean-field configurations. The resulting
wave functions encode more correlations than the ones obtained within
SR methods while still keeping well defined symmetry quantum
numbers. \cite{Tomita-1}

There are differents flavors of MR approximations available in the
literature.\cite{Carlo-review,Fukutome-original-RSHF,Tomita-1,Yamamoto-1,Yamamoto-2,Ikawa-1993,Tomita-2,Tomita-3,Mizusaki-1,Mizusaki-2}
In the present study, we adopt a MR scheme well known in nuclear
structure physics, \cite{Carlo-review} which to the best of our
knowledge has not been applied to lattice models. The key ingredient
in such MR scheme is the inclusion of relevant correlations in both
ground and excited states on an equal footing. As a benchmark test, we
concentrate on the 1D Hubbard model for which exact solutions are
known. \cite{LIEB,BETHE} In particular, we consider the case of
half-filled lattices. Nevertheless, the present MR approximation can
be extended to the 2D case as well as to doped systems with arbitrary
on-site interaction strengths.

For a given single-electron space we resort to generalized
HF-transformations \cite{StuberPaldus} (GHF) mixing all quantum
numbers of the single-electron basis states. The corresponding Slater
determinants deliberately break spin and spatial symmetries of the 1D
Hubbard Hamiltonian. \cite{text-Hubbard-1D} We restore these broken
symmetries with the help of projection operators. \cite{rs} The
resulting MR ground state wave functions are obtained applying the
variational principle to the projected energy.

The structure of our MR ground state wave functions is formally
similar to the one adopted within the Resonating HF \cite
{Fukutome-original-RSHF,Yamamoto-1,Yamamoto-2,Ikawa-1993,Tomita-1,Tomita-2}
(ResHF) method, i.e., they are expanded in terms of a given number of
non-orthogonal symmetry-projected configurations. Nevertheless, while
in the latter all the underlying HF-transformations and mixing
coefficients are optimized simultaneously, \cite{Tomita-1,Tomita-2} in
our case the orbital optimization is performed sequentially, only for
the last added HF-transformation (all our mixing coefficients are
still optimized at the same time) rendering our calculations easier to
handle. This is particularly relevant for alleviating our numerical
effort if one keeps in mind that, for both ground and excited states,
we use the most general GHF-transformations and therefore a full 3D
spin projection is required.

Our MR scheme is also used to compute spin-spin correlation functions
(SSCFs) in real space, magnetic structure factors (MSFs) as well as
dynamical properties of the 1D Hubbard model like spectral functions
(SFs) and density of states (DOS). \cite%
{Dagotto-review,Rayner-2D-Hubbard-PRB-2012,Carlos-Hubbard-1D,Fetter-W}
On the other hand, one may wonder whether there is any relevant
information in the intrinsic symmetry-broken GHF-determinants
associated with our MR wave functions. As will be shown below, the
structure of such intrinsic determinants can be interpreted in terms
of basic units of quantum fluctuations for the lattices
considered.\cite{Horovitz}

In addition to ground state properties, our MR framework treats
excited states with well defined quantum numbers as expansions in
terms of non-orthogonal symmetry-projected configurations using chains
of variation-after-projection (VAP) calculations. As a byproduct, we
also obtain a (truncated) basis consisting of a few Gram-Schmidt
orthonormalized states, \cite{Rayner-2D-Hubbard-PRB-2012} which may be
used to perform a final diagonalization of the Hamiltonian in order to
account for further correlations in both ground and excited states.

The layout of the theory part of this paper is as follows. First, we
introduce the methodology of our MR VAP scheme in
Sec.\ref{Theory}. Symmetry restoration based on a single Slater
determinant (i.e., SR symmetry restoration) is described in
Sec.\ref{Hubbard1DHamiltonian}. This section will serve to set our
notation as well as to introduce some key elements of our 3D spin and
full space group projection techniques. Subsequently, symmetry
restoration based on several Slater determinants (i.e., MR symmetry
restoration) is discussed in Sec.\ref{Symmetry-multiref}. In
particular, the MR description of ground and excited states is
presented in Secs.\ref{Symmetry-multiref-gs} and
\ref{Symmetry-multiref-excited}, respectively. In
Sec.{\ref{SFplusDOS}}, we will briefly discuss the computation of the
SFs and DOS within our theoretical framework.

The results presented in this paper test the performance of our
approximation in a selected set of illustrative examples. In most
cases, calculations have been carried out for on-site repulsions
$U=2t,4t$ and $8t$ taken as representatives of weak,
intermediate-to-strong, and strong correlation regimes. In
Sec.\ref{Results}, we first consider the ground states of half-filled
lattices with up to 50 sites. We compare our ground state and
correlation energies with the exact ones as well as with those
obtained using other theoretical methods. We then discuss the
dependence of the predicted correlation energies on the number of
non-orthogonal symmetry-projected configurations used to expand our
ground state wave functions, the computational performance of our
scheme as well as the structure of the intrinsic GHF-determinants
resulting from our MR VAP procedure. Next, we consider the results of
our calculations for SSCFs in real space and MSFs for half-filled
lattices with up to 30 sites. These results are compared with 
density matrix renormalization
group \cite{DMRG-White,Dukelsky-Pittel-RPP,Scholl-RMP,Xiang}
(DMRG) 
ones obtained with the open source ALPS software.\cite{ALPS} 
This comparison is very valuable as DMRG represents one of the 
most accurate approximations in the  1D case. Subsequently, we compare the DOS
provided by our theoretical framework with the exact one, obtained
with an in-house diagonalization code, in a lattice with 10
sites. Results for hole SFs are also discussed for a 30-site
lattice. We end Sec.\ref{Results} by presenting results for excitation
spectra in various lattices and discussing the structure of the
intrinsic GHF-determinants resulting from our MR VAP procedure for
excited states. Finally, Sec.\ref{conclusions} is devoted to
concluding remarks and work perspectives.

\begin{table}[tbp]
\caption{Ground state energy of the half-filled  lattices with $N_{sites}=12$ and $20$, as predicted with the GHF-FED scheme based on $n_{1}=10$
GHF-transformations, are compared with exact results for on-site repulsions of $U=2t,4t$ and $8t$. Energies obtained
with the RHF and UHF approximations are included as a reference.
The ratio of correlation energies $\kappa$ obtained with the UHF
and GHF-FED aproximations, is computed according to Eq.(\ref{formulaCE}). For more details, see the main text. }
\label{Table1}
\begin{tabular}{cccccccccccc}
\hline\hline
&  & $N_{sites}=12$ &  & $\kappa(\%)$ &  & $N_{sites}=20$ &  & $\kappa(\%)$
&  &  &  \\ \hline
&  &  &  &  &  &  &  &  &  &  &  \\
U=2t & RHF & -8.9282 &  &  &  & -15.2551 &  &  &  &  &  \\
& UHF & -9.3379 &  & 36.79 &  & -15.6411 &  & 24.05 &  &  &  \\
& GHF-FED & -10.0401 &  & 99.85 &  & -16.8565 &  & 99.79 &  &  &  \\
& EXACT & -10.0418 &  &  &  & -16.8599 &  &  &  &  &  \\
&  &  &  &  &  &  &  &  &  &  &  \\
&  &  &  &  &  &  &  &  &  &  &  \\
U=4t & RHF & -2.9282 &  &  &  & -5.2550 &  &  &  &  &  \\
& UHF & -5.6290 &  & 67.65 &  & -9.3821 &  & 66.08 &  &  &  \\
& GHF-FED & -6.9201 &  & 99.99 &  & -11.4954 &  & 99.92 &  &  &  \\
& EXACT & -6.9204 &  &  &  & -11.5005 &  &  &  &  &  \\
&  &  &  &  &  &  &  &  &  &  &  \\
&  &  &  &  &  &  &  &  &  &  &  \\
U=8t & RHF & 9.0718 &  &  &  & 14.7450 &  &  &  &  &  \\
& UHF & -2.9532 &  & 92.26 &  & -4.9219 &  & 92.23 &  &  &  \\
& GHF-FED & -3.9625 &  & 99.99 &  & -6.5612 &  & 99.96 &  &  &  \\
& EXACT & -3.9626 &  &  &  & -6.5699 &  &  &  &  &  \\
&  &  &  &  &  &  &  &  &  &  &  \\ \hline
\end{tabular}
\end{table}

\section{Theoretical Framework}

\label{Theory} In what follows, we describe the theoretical framework used
in the present study. First, SR symmetry restoration is presented in
Sec.\ref{Hubbard1DHamiltonian}. Subsequently, in
Sec.\ref{Symmetry-multiref}, we consider our MR scheme to describe
both ground (Sec.\ref{Symmetry-multiref-gs}) and excited
(Sec.\ref{Symmetry-multiref-excited}) states of the 1D Hubbard
model. The computation of SFs and DOS is briefly discussed in
Sec.{\ref{SFplusDOS}}.

\subsection{Single-reference (SR) symmetry restoration}

\label{Hubbard1DHamiltonian}

We consider the 1D Hubbard Hamiltonian \cite{Hubbard-model_def1}

\begin{equation} \label{HAM-hubbard1D}
\hat{H}=-t\sum_{j,\sigma }\Big \{\hat{c}_{j+1\sigma }^{\dagger }\hat{c}_{j\sigma }+\hat{c}_{j\sigma }^{\dagger }\hat{c}_{j+1\sigma }\Big \}%
+U\sum_{j}\hat{n}_{j\uparrow }\hat{n}_{j\downarrow }
\end{equation}
where the first term represents the nearest-neighbor hopping ($t>0$)
and the second is the repulsive on-site interaction ($U>0$). The
fermionic \cite{Blaizot-Ripka} operators $\hat{c}_{j\sigma }^{\dagger
}$ and $\hat{c}_{j\sigma }$ create and destroy an electron with
spin-projection $\sigma =\pm 1/2$ (also denoted as $\sigma =\uparrow
,\downarrow $) along an arbitrary chosen quantization axis on a
lattice site $j=1,\dots ,N_{sites}$.  The operators $\hat{n}_{j\sigma
}=\hat{c}_{j\sigma }^{\dagger }\hat{c}_{j\sigma }$ are the local
number operators. We assume periodic boundary conditions, i.e., the
sites $j$ and $j+N_{sites}$ are identical.  Furthermore, we assume a
lattice spacing $\Delta =1$.

In the standard HF-approximation, \cite{Blaizot-Ripka,rs} the ground
state of an $N_{e}$-electron system is represented by a Slater
determinant $|{ \mathcal{D}}\rangle
=\prod_{h=1}^{N_{e}}\hat{b}_{h}^{+}|0\rangle $ in which the
energetically lowest $N_{e}$ single-fermion states (holes
$h,h^{^{\prime }},\dots $) are occupied while the remaining
$2N_{sites}-N_{e}$ states (particles $p,p^{^{\prime }},\dots $) are
empty. For a set of single-fermion operators $\hat{c}^{\dagger }$, the
HF-quasiparticle operators $\hat{b} ^{\dagger }$ are given by the
following canonical transformation \cite{Blaizot-Ripka,rs}

\begin{equation} \label{HF-transformation}
\hat{b}_{i}^{\dagger }=\sum_{j\sigma }{\mathcal{D}}_{j\sigma ,i}^{\ast }\hat{%
c}_{j\sigma }^{\dagger }
\end{equation}
where ${\mathcal{D}}$ is a general $2N_{sites}\times 2N_{sites}$ unitary
\cite{non-unitary-paper-Carlos} 
matrix, i.e., ${\mathcal{D}}{\mathcal{D}}^{\dagger }={\mathcal{D}}^{\dagger }{\mathcal{D}}=1$. In all the
calculations to be discussed below, we have used generalized HF (GHF)
transformations. \cite{StuberPaldus} As it is well known, the most general
GHF-determinant $|{\mathcal{D}}\rangle $ deliberately breaks several
symmetries of the original Hamiltonian. \cite%
{Carlo-review,rs,Rayner-2D-Hubbard-PRB-2012,Carlos-Hubbard-1D,PQT-reference-1,PQT-reference-2}
Typical examples are the rotational (in spin space) and spatial symmetries.
To restore the spin quantum numbers in a symmetry-broken GHF-determinant, we
explicitly use the full 3D projection operator 
\cite{Rayner-2D-Hubbard-PRB-2012,Carlos-Hubbard-1D,PQT-reference-1,PQT-reference-2}

\begin{equation} \label{PROJ-S} 
\hat{P}_{\Sigma {\Sigma }^{^{\prime }}}^{S}=\frac{2S+1}{8{\pi }^{2}}\int
d\Omega {\mathcal{D}}_{\Sigma {\Sigma }^{^{\prime }}}^{S\ast }(\Omega)R(\Omega )
\end{equation}
where $R(\Omega )=e^{-i\alpha \hat{S}_{z}}e^{-i\beta \hat{S}_{y}}e^{-i\gamma
\hat{S}_{z}}$ is the rotation operator in spin space, the label $\Omega
=\left( \alpha ,\beta ,\gamma \right) $ stands for the set of Euler angles
and ${\mathcal{D}}_{\Sigma {\Sigma }^{^{\prime }}}^{S}(\Omega )$ are Wigner
functions. \cite{Edmonds} To recover the spatial symmetries, we introduce
the projection operator

\begin{equation} \label{proj-space-group}
\hat{P}_{mm^{^{\prime }}}^{k}=\frac{1}{2N_{sites}}\sum_{g}{\Gamma }%
_{mm^{^{\prime }}}^{k}(g)\hat{R}(g)
\end{equation}
where ${\Gamma }_{mm^{^{\prime }}}^{k}(g)$ is the matrix
representation of an irreducible representation, which can be found by
standard methods, \cite{Tomita-1,Lanczos-Fano} and $ \hat{R}(g)$
represents the corresponding symmetry operations (i.e., translation by
one lattice site and the reflection $x\rightarrow -x$) parametrized in
terms of the label $g$. The linear momentum ${k}=(2\pi / N_{sites}) \,
{\xi }$ is given in terms of the quantum number $\xi $ that takes the
values

\begin{table}[tbp]
\caption{Ground state energy of the half-filled  lattices with $N_{sites}=30$ and $50$
predicted with the GHF-FED scheme based on $n_{1}=25$ GHF-transformations,
are compared with exact results for on-site
repulsions of $U=2t,4t$ and $8t$. Results obtained with the UHF-ResHF
approximation, \cite{Tomita-1} based on $n_{1}=30$
UHF-transformations, as well as the RHF and UHF energies are also included
in the table. The ratio of correlation energies $\kappa $ obtained
with the UHF, UHF-ResHF and the GHF-FED aproximations, is computed according
to Eq.(\ref{formulaCE}).}
\label{Table2}
\begin{tabular}{cccccccccccc}
\hline\hline
&  & $N_{sites}=30$ &  & $\kappa(\%)$ &  & $N_{sites}=50$ &  & $\kappa(\%)$
&  &  &  \\ \hline
&  &  &  &  &  &  &  &  &  &  &  \\
U=2t & RHF & -23.2671 &  &  &  & -38.7039 &  &  &  &  &  \\
& UHF & -23.4792 &  & 10.02 &  & -39.1294 &  & 12.02 &  &  &  \\
& UHF-ResHF & -25.3436 &  & 98.11 &  & -41.9535 &  & 91.78 &  &  &  \\
& UHF-FED & -25.3508 &  & 98.45 &  & -41.9963 &  & 92.99 &  &  &  \\
& GHF-FED & -25.3730 &  & 99.50 &  & -42.1219 &  & 96.46 &  &  &  \\
& EXACT & -25.3835 &  &  &  & -42.2443 &  &  &  &  &  \\
&  &  &  &  &  &  &  &  &  &  &  \\
&  &  &  &  &  &  &  &  &  &  &  \\
U=4t & RHF & -8.2671 &  &  &  & -13.7039 &  &  &  &  &  \\
& UHF & -14.0732 &  & 64.75 &  & -23.4553 &  & 65.02 &  &  &  \\
& UHF-ResHF & -17.0542 &  & 98.00 &  & -27.9633 &  & 95.09 &  &  &  \\
& UHF-FED & -16.9420 &  & 96.75 &  & -27.3518 &  & 91.01 &  &  &  \\
& GHF-FED & -17.1789 &  & 99.39 &  & -27.9788 &  & 95.19 &  &  &  \\
& EXACT & -17.2335 &  &  &  & -28.6993 &  &  &  &  &  \\
&  &  &  &  &  &  &  &  &  &  &  \\
&  &  &  &  &  &  &  &  &  &  &  \\
U=8t & RHF & 21.7329 &  &  &  & 36.2961 &  &  &  &  &  \\
& UHF & -7.8329 &  & 93.65 &  & -12.3048 &  & 92.26 &  &  &  \\
& UHF-ResHF & -9.5378 &  & 99.04 &  & -15.6422 &  & 98.59 &  &  &  \\
& UHF-FED & -9.3524 &  & 98.46 &  & -14.8461 &  & 97.08 &  &  &  \\
& GHF-FED & -9.7612 &  & 99.75 &  & -15.6753 &  & 98.65 &  &  &  \\
& EXACT & -9.8387 &  &  &  & -16.3842 &  &  &  &  &  \\
&  &  &  &  &  &  &  &  &  &  &  \\ \hline
\end{tabular}
\end{table}

\begin{equation} \label{values-momenta}
{\xi }=-\frac{N_{sites}}{2}+1,\dots ,\frac{N_{sites}}{2}
\end{equation}
allowed inside the Brillouin zone (BZ). \cite{Ashcroft-Mermin-book}
Equivalently, it can take all integer values between 0 and $N_{sites}-1$.
For $k=0,\pi $ an additional label $b=\pm 1$ should be introduced to account
for the parity of the corresponding irreducible representation under the
reflection $x\rightarrow -x$. \cite{Tomita-1,Lanczos-Fano} In what follows,
we do not explictly write this label $b$ but the reader should keep in mind
that it is taken into account whenever needed.

We introduce \cite{Rayner-2D-Hubbard-PRB-2012} the shorthand notation $%
\Theta =(S,k)$ for the set of symmetry (i.e., spin and linear momentum)
quantum numbers as well as $K=(\Sigma ,m)$. The total projection operator
reads

\begin{eqnarray}  \label{projection-operator-P}
\hat{P}_{K K^{^{\prime }}}^{\Theta} \equiv \hat{P}_{\Sigma
  {\Sigma}^{^{\prime }} }^{S } \hat{P}_{m m^{^{\prime }}}^{k}
\end{eqnarray}

We then superpose the Goldstone manifold 
$\hat{R}(\Omega) \hat{R}(g)| {\mathcal{D}} \rangle$ to recover the spin and spatial symmetries \cite{rs}
via the following SR ansatz

\begin{eqnarray}  \label{SP-WF-VAMPIR}
| {\mathcal{D}}; \Theta; K \rangle = \sum_{K^{^{\prime }}} f_{K^{^{\prime }}
}^{ \Theta } \hat{P}_{K K^{^{\prime }}}^{\Theta} | {\mathcal{D}} \rangle
\end{eqnarray}
where $f_{}^{ \Theta }$ are variational parameters. Note, that the state Eq.(%
\ref{SP-WF-VAMPIR}) is already multi-determinantal \cite%
{Rayner-2D-Hubbard-PRB-2012,PQT-reference-1} via the projection operator $%
\hat{P}_{K K^{^{\prime }}}^{\Theta}$. For a given symmetry $\Theta$, the
energy (independent of K) associated with the state Eq.(\ref{SP-WF-VAMPIR})

\begin{eqnarray}  \label{energy-vampir}
E^{\Theta} = \frac{ f^{ \Theta \dagger} {\mathcal{H}}^{ \Theta } f^{ \Theta
} } { f^{ \Theta \dagger} {\mathcal{N}}^{ \Theta } f^{ \Theta }}
\end{eqnarray}
is given in terms of the Hamiltonian and norm

\begin{eqnarray}  \label{HNkernels-GS-SR}
{\mathcal{H}}_{K K^{^{\prime }}}^{ \Theta } &=& \langle {\mathcal{D}} 
| \hat{H} \hat{P}_{K K^{^{\prime }}}^{\Theta} | {\mathcal{D}} \rangle  
\nonumber\\
{\mathcal{N}}_{K K^{^{\prime }}}^{ \Theta } &=& \langle {\mathcal{D}} | \hat{
P}_{K K^{^{\prime }}}^{\Theta} | {\mathcal{D}} \rangle
\end{eqnarray}
matrices. It has to be minimized with respect to the coefficients $f^{\Theta}$ and the underlying GHF-transformation ${\mathcal{D}}$. The
variation with respect to the former yields the following resonon-like \cite{Gutzwiller_method} eigenvalue equation \cite{Rayner-2D-Hubbard-PRB-2012,Carlos-Hubbard-1D}

\begin{eqnarray}  \label{HW-1}
\left({\mathcal{H}}^{\Theta} - E^{ \Theta} {\mathcal{N}}^{\Theta} \right)
f^{\Theta} = 0
\end{eqnarray}
with the constraint $f^{\Theta \dagger} {\mathcal{N}}^{\Theta } f^{\Theta }
= 1$ ensuring the orthogonality of the solutions. On the other hand, the
unrestricted minimization of the energy [Eq.(\ref{energy-vampir})] with
respect to ${\mathcal{D}}$ is carried out via the Thouless theorem. \cite{Rayner-2D-Hubbard-PRB-2012,Carlos-Hubbard-1D,non-unitary-paper-Carlos}

For a given symmetry $\Theta $, we only retain the energetically lowest
solution of our VAP equations. \cite{Rayner-2D-Hubbard-PRB-2012} Both the GHF-transformation ${\mathcal{D}}$ and
the mixing coefficients $f^{\Theta }$ are complex, therefore one needs to
minimize $n_{var}=2(2N_{sites}-N_{e})\times N_{e}+4S$ real variables. We use
a limited-memory quasi-Newton method for such minimization. 
\cite{Rayner-2D-Hubbard-PRB-2012,Carlos-Hubbard-1D,quasi-Newton-3}
In practice, the integration over the set of Euler angles in Eq.(\ref{PROJ-S}) is 
discretized. For example, for a lattice with $N_{sites}=30$ we have used
13, 26, and 13 grid
points for the integrations over $\alpha$, $\beta$, and $\gamma$, respectively.
In this case, a total of 263,640 grid points are used in the
discretization of the projection operator of Eq.(\ref{projection-operator-P}). We
have afforded such a task by developing a parallel implementation for all
the VAP  schemes discussed in this paper.

\subsection{Multi-reference (MR) symmetry restoration}

\label{Symmetry-multiref}

For each symmetry $\Theta $, the SR procedure described in 
Sec.\ref{Hubbard1DHamiltonian} provides us with the optimal 
variational representation of the corresponding ground state via a single
symmetry-projected
GHF-determinant. However, as the lattice size increases one may adopt a MR
perspective to keep and/or improve the quality of the wave functions. 
\cite{Carlo-review,Fukutome-original-RSHF} The key features of our MR approach,
known in nuclear structure physics as the FED VAMP \cite{Carlo-review} (Few
Determinant Variation After Mean-field Projection) strategy, for the considered
ground states are described in the next subsection. We
use the acronym GHF-FED to refer to it in the present work. On the other
hand, our MR approach for excited states, known 
as EXCITED FED VAMP, \cite{Carlo-review} will be presented in Sec.\ref{Symmetry-multiref-excited}. We
will use the acronym GHF-EXC-FED to refer to it in what follows.

%
%
\begin{figure}[tbp]
\includegraphics[width=0.45\textwidth]{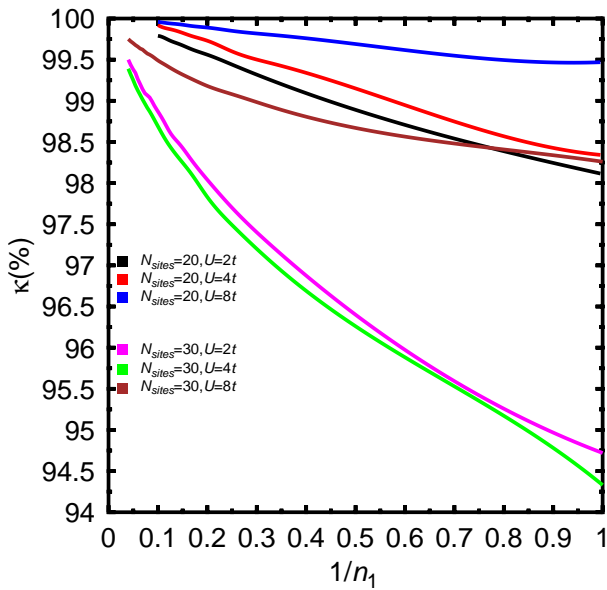}
\caption{(Color online) The ratio of correlation energies $\kappa $
obtained with the GHF-FED approximation is
plotted as a function of the inverse of the number of GHF-transformations
for the half-filled  lattices
with $N_{sites}=20$ and $30$. Results are shown for on-site repulsions of $U=2t,4t$ and $8t$. For more details, see the main text. }
\label{energy_vs_trans}
\end{figure}
%
%

\subsubsection{MR symmetry restoration for ground states (GHF-FED)}
\label{Symmetry-multiref-gs}

 Our goal in this section is to obtain, through a chain of VAP
 calculations, non-orthogonal symmetry-projected GHF-configurations
 used to build a MR expansion of a given ground state
 \cite{Carlo-review} with well defined symmetry quantum numbers
 $\Theta$.

Suppose we have generated a ground state solution $|\phi _{1K}^{1\Theta
}\rangle =|{\mathcal{D}}_{1}^{1};\Theta ;K\rangle $ [Eq.(\ref{SP-WF-VAMPIR}%
)]. Note that at this point, we have added the superscript $1$ to explicitly
indicate that only one GHF-transformation has been used within the SR
approximation discussed in Sec.\ref{Hubbard1DHamiltonian}. On the other
hand, the subscript $1$ has been added to indicate that the ground state is
considered. As we will see in Sec.\ref{Symmetry-multiref-excited},
this subscript will allow us to distinguish between ground (i.e., $i=1$) and
excited (i.e., $i=2,3,\dots ,m$) states. On the other hand, both indices are
also added to the (intrinsic) GHF-transformation to explicitly indicate that
it is variationally optimized for the state $|\phi _{1K}^{1\Theta }\rangle $%
. We then keep the transformation ${\mathcal{D}}_{1}^{1}$ fixed and consider
the ansatz

\begin{equation} \label{FED-vampir-example1}
|\phi _{1K}^{2\Theta }\rangle =\sum_{K^{^{\prime
}}}\sum_{i=1}^{2}f_{1K^{^{\prime }}}^{i\Theta }\hat{P}_{KK^{^{\prime
}}}^{\Theta }|{\mathcal{D}}_{1}^{i}\rangle
\end{equation}
which approximates the ground state (subscript 1) by means of two
(superscript 2) non-orthogonal symmetry-projected GHF-determinants. It
is obtained applying the variational principle to the energy
functional with respect to the last added transformation
${\mathcal{D}}_{1}^{2}$ and all the new mixing coefficients
$f_{1}^{i\Theta }$. A similar procedure can be followed to approximate
the ground state by a larger number of non-orthogonal
symmetry-projected configurations. Let us assume that $n_{1}-1$
configurations have already been computed. Then, one introduces a new
GHF-transformation ${\mathcal{D}}_{1}^{n_{1}}$, a new set of mixing
coefficients $f_{1}^{i\Theta },$ and makes the MR GHF-FED ansatz

\begin{eqnarray}  \label{FED-state-general}
| \phi_{1 K}^{n_{1} \Theta} \rangle = \sum_{K^{^{\prime }}}
\sum_{i=1}^{n_{1}} f_{1 K^{^{\prime }} }^{i \Theta} \hat{P}_{K K^{^{\prime
}}}^{\Theta} | {\mathcal{D}}_{1}^{i} \rangle
\end{eqnarray}
which superposes the Goldstone manifolds $\hat{R}(\Omega) \hat{R}(g) | {\mathcal{D}}_{1}^{i} \rangle$. 
The corresponding energy

\begin{eqnarray}  \label{ojo-ojo}
E_{1}^{n_{1} \Theta} = \frac{ f^{n_{1} \Theta \dagger} {\mathcal{H}}^{n_{1}
\Theta} f^{n_{1} \Theta} } { f^{n_{1} \Theta \dagger} {\mathcal{N}}^{n_{1}
\Theta} f^{n_{1} \Theta} }
\end{eqnarray}
is given in terms of the Hamiltonian and norm

\begin{eqnarray} \label{HNKernels-GHF-FED}
{\mathcal{H}}_{iK,jK^{^{\prime }}}^{n_{1}\Theta } &=&
\langle {\mathcal{D}}_{1}^{i}|\hat{H}\hat{P}_{KK^{^{\prime }}}^{\Theta }|{\mathcal{D}}_{1}^{j}\rangle  
 \nonumber\\
{\mathcal{N}}_{iK,jK^{^{\prime }}}^{n_{1}\Theta } &=&\langle {\mathcal{D}}_{1}^{i}|\hat{P}_{KK^{^{\prime }}}^{\Theta }|{\mathcal{D}}_{1}^{j}\rangle
\end{eqnarray}
kernels, which require the knowledge of the symmetry-projected matrix elements between all
the GHF-determinants used in the expansion Eq.(\ref{FED-state-general}). The
wave function Eq.(\ref{FED-state-general}) is determined varying the energy
Eq.(\ref{ojo-ojo}) with respect to all the new mixing coefficients 
$f_{1}^{i\Theta }$ and the last added transformation ${\mathcal{D}}_{1}^{n_{1}}$. In the former case, we obtain an eigenvalue equation similar
to Eq.(\ref{HW-1}), with the constraint 
$f^{n_{1}\Theta \dagger }{\mathcal{N}}^{n_{1}\Theta }f^{n_{1}\Theta }=1$, while the unrestricted minimization
with respect to ${\mathcal{D}}_{1}^{n_{1}}$ is carried out via the Thouless
theorem. Let
us stress that the GHF-FED MR approximation Eq.(\ref{FED-state-general}) of
a given ground state enlarges the flexibity in our wave functions to a
total number of $n_{var}=2n_{1}(2N_{sites}-N_{e})\times N_{e}+4n_{1}S+2(n_{1}-1)$ variational parameters.

%
%
\begin{figure}[tbp]
\includegraphics[width=0.45\textwidth]{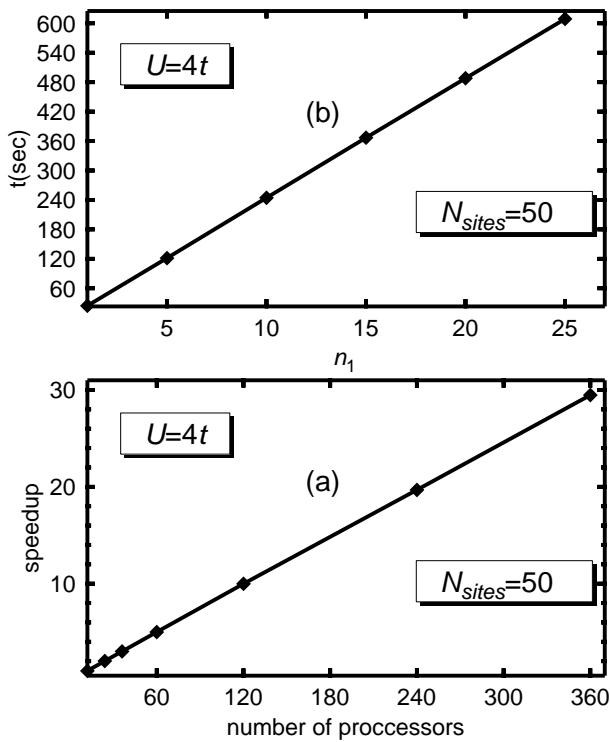}
\caption{Speedup of a typical GHF-FED calculation is shown in panel (a) as a
function of the number of proccessors. The corresponding scaling (for a
fixed number of proccessors) with the number of transformations $n_{1}$ is presented in panel
(b). Results are for the half-filled  lattice with $N_{sites}=50$ and $U=4t.$}
\label{sppedupplot}
\end{figure}
%
%

\subsubsection{MR symmetry restoration for excited states (GHF-EXC-FED)}

\label{Symmetry-multiref-excited} In this section, we construct
non-orthogonal symmetry-projected GHF-configurations to expand a given excited state. The
orthogonalization between ground and excited states is achieved via the
Gram-Schmidt procedure. \cite{Rayner-2D-Hubbard-PRB-2012} As a byproduct,
our MR GHF-EXC-FED method also yields a (truncated) basis consisting of a
few orthonormal states which may be used to diagonalize the
Hamiltonian and account for further correlations in both ground and excited
states. \cite{Rayner-2D-Hubbard-PRB-2012,Carlo-review}

Let us assume that we have already obtained a GHF-FED ground 
state $|\phi_{1}^{n_{1}}\rangle =|\phi _{1K}^{n_{1}\Theta }\rangle $ [Eq.(\ref%
{FED-state-general})] along the lines discussed in the previous Sec.\ref%
{Symmetry-multiref-gs}. We then look for the first excited state (subscript
2) with the same symmetry $\Theta $, approximated by a given $n_{2}$ number
of non-orthogonal symmetry-projected  configurations. We start with the ansatz

\begin{equation} \label{FED-excited-example1}
|\varphi _{2}^{1}\rangle =\alpha ^{1}|\phi _{1}^{n_{1}}\rangle +\beta^{1}|\phi _{2}^{1}\rangle
\end{equation}
where $|\phi _{2K}^{1\Theta }\rangle $ has a form similar to Eq.(\ref{SP-WF-VAMPIR}) but
 written in terms of the coefficients $f_{2}^{1\Theta }$
and the GHF-determinant $|{\mathcal{D}}_{2}^{1}\rangle $. Both $\alpha ^{1}$
and $\beta ^{1}$ can be obtained by requiring orthonormalization with
respect to the ground state that we already have.
The state Eq.(\ref{FED-excited-example1}) is
determined varying the energy functional with respect to $f_{2}^{1\Theta }$
and ${\mathcal{D}}_{2}^{1}$. When $n_{2}-1$ configurations have already been
computed for the first excited state, one makes the ansatz

\begin{eqnarray}  \label{FED-excited-example-coco}
| \varphi_{2}^{n_{2}} \rangle = \alpha^{n_{2}} | \phi_{1}^{n_{1}} \rangle +
\beta^{n_{2}} | \phi_{2}^{n_{2}} \rangle
\end{eqnarray}
where the state $| \phi_{2 K}^{n_{2} \Theta} \rangle$ has a form similar to
Eq.(\ref{FED-state-general}) but written in terms of the new coefficients $%
f_{2}^{i \Theta}$ and the GHF-transformations ${\mathcal{D}}_{2}^{i}$ (i= 1,
$\dots$, $n_{2}$). Once again, the coefficients $\alpha^{n_{2}}
$ and $\beta^{n_{2}}$ are obtained by requiring orthonormalization with
respect to the ground state we already have. The wave function Eq.(\ref%
{FED-excited-example-coco}) is determined varying the energy functional with
respect to the last added GHF-transformation ${\mathcal{D}}_{2}^{n_{2}}$ and
all the coefficients $f_{2}^{i \Theta}$.

Now, we consider the most general situation in which the ground state $%
|\varphi _{1}^{n_{1}}\rangle =|\phi _{1}^{n_{1}}\rangle $ as well as a set
of $m-2$ Gram-Schmidt orthonormalized excited states $|\varphi
_{2}^{n_{2}}\rangle $, $|\varphi _{3}^{n_{3}}\rangle $, $\dots $, $|\varphi
_{m-1}^{n_{m-1}}\rangle $, all of them with the same symmetry quantum
numbers $\Theta $, are already at our disposal. Each of these $m-1$ states
is optimized using chains of VAP calculations, as discussed in Sec.\ref%
{Symmetry-multiref-gs} and in the present section. The key question is then
how to approximate the $m^{th}$ excited state by $n_{m}$ non-orthogonal
symmetry-projected configurations. We also need to ensure orthogonality with respect to
all the $m-1$ states we already have. Let us assume that $n_{m}-1$
configurations have already been computed for the $m^{th}$ excited state
with symmetry $\Theta $. Then an approximation in terms of $n_{m}$
non-orthogonal symmetry-projected GHF-configurations is obtained with the GHF-EXC-FED ansatz

\begin{equation} \label{FED-excited-example-ansatz2}
|\varphi _{m}^{n_{m}}\rangle =\sum_{i=1}^{m-1}\omega _{i}^{m}|\phi
_{i}^{n_{i}}\rangle +\tau ^{m}|\phi _{m}^{n_{m}}\rangle
\end{equation}
where the state $|\phi _{mK}^{n_{m}\Theta }\rangle $ has a form similar to
Eq.(\ref{FED-state-general}) but is written in terms of new coefficients $%
f_{m}^{i\Theta }$ and GHF-transformations ${\mathcal{D}}_{m}^{i}$ (i= 1, $%
\dots $, $n_{m}$). The coefficients $\tau ^{m}$ and $\omega
_{i}^{m}$ read

\begin{eqnarray}
\tau^{m} &=& \langle \phi_{m }^{n_{m}} | \left(1- \hat{S}_{m-1} \right) |
\phi_{m }^{n_{m}} \rangle^{-1/2}  \notag \\
\omega_{i}^{m} &=& - \sum_{k=1}^{m-1} \left({\mathcal{A}}^{-1} \right)_{ik}
\langle \phi_{k }^{n_{k}} | \phi_{m }^{n_{m}}\rangle \tau^{m}
\end{eqnarray}
in terms of the projector (i.e., $\hat{S}_{m-1} = \hat{S}_{m-1}^{2}$)

\begin{equation} \label{mminus1projectorS}
\hat{S}_{m-1}=\sum_{i,k=1}^{m-1}|\phi _{i}^{n_{i}}\rangle \left( {\mathcal{A}%
}^{-1}\right) _{ik}\langle \phi _{k}^{n_{k}}|
\end{equation}
with the overlap matrix ${\mathcal{A}}_{ik}=\langle \phi _{i}^{n_{i}}|\phi
_{k}^{n_{k}}\rangle $. The MR GHF-EXC-FED wave function Eq.(\ref%
{FED-excited-example-ansatz2}) is determined by varying all the coefficients
$f_{m}^{i\Theta }$ and the last added GHF-transformation ${\mathcal{D}}_{m}^{n_{m}}$. The energy is

%
%
\begin{figure*}[tbp]
\includegraphics[width=1.0\textwidth]{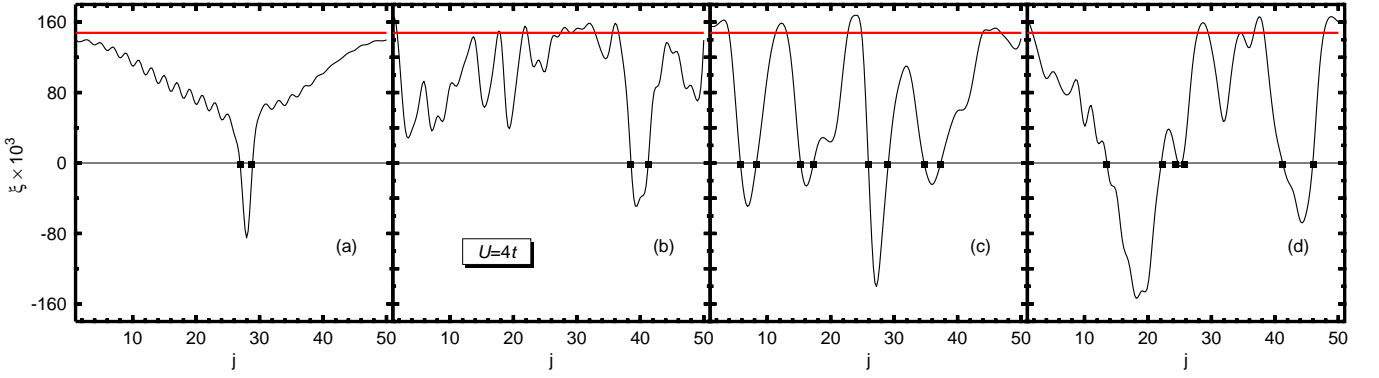}
\caption{(Color online) The quantity ${\protect\xi }_{1}^{i}$ [Eq.(\protect
\ref{SDW1})] is plotted as a function of lattice site $j$ for some typical
symmetry-broken GHF-determinants $|D_{1}^{i}\rangle $ resulting from the
GHF-FED VAP  optimization for the half-filled  lattice with $N_{sites}=50$ and $U=4t$. The UHF 
spin-density wave
 is plotted in red for comparison. For
more details, see the main text.}
\label{solitonsU2}
\end{figure*}
%
%

\begin{equation} \label{ojo-ojo-excitedvamp}
E_{m}^{n_{m}\Theta }=\frac{f^{n_{m}\Theta \dagger }
{\mathcal{H}}^{n_{m}\Theta }f^{n_{m}\Theta }}{f^{n_{m}\Theta \dagger }{\mathcal{N}}^{n_{m}\Theta }f^{n_{m}\Theta }}
\end{equation}
where the Hamiltonian ${\mathcal{H}}^{n_{m}\Theta }$ and 
norm ${\mathcal{N}}^{n_{m}\Theta }$ kernels account for the fact that $m-1$ linearly
independent solutions have been removed from the variational 
space.
The kernel expressions are slightly more
involved than the ones in Eqs.(\ref{HNkernels-GS-SR}) and (\ref{HNKernels-GHF-FED}) but
 still can be obtained straightforwardly. The
variation with respect to the coefficients $f_{m}^{i\Theta }$ leads to a
generalized eigenvalue equation similar to Eq.(\ref{HW-1}) with the
constraint $f^{n_{m}\Theta \dagger }{\mathcal{N}}^{n_{m}\Theta}f^{n_{m}\Theta }=1$, while the 
unrestricted minimization with respect to
the last added GHF-transformation ${\mathcal{D}}_{m}^{n_{m}}$ is carried out
via the Thouless theorem.

The GHF-EXC-FED scheme outlined in this section
provides, for each set of symmetry quantum numbers $\Theta $, a (truncated)
basis of $m$ (orthonormalized) states $|\varphi _{1K}^{n_{1}\Theta }\rangle,\dots ,|\varphi _{mK}^{n_{m}\Theta }\rangle $, each 
of them expanded by $n_{1}$, $\dots $,$n_{m}$ non-orthogonal symmetry-projected GHF-determinants, respectively.
Finally, the diagonalization of the Hamiltonian Eq.(\ref{HAM-hubbard1D}) in such a basis

\begin{eqnarray}  \label{diagonalizationfinal}
\sum_{j=1}^{m} \Big[ \langle \varphi_{i}^{n_{i}} | \hat{H} |
\varphi_{j}^{n_{j}} \rangle - {\epsilon}_{\alpha}^{\Theta} {\delta}_{ij} \Big] C_{j \alpha}^{\Theta} = 0
\end{eqnarray}
provides ground and excited states

\begin{equation} \label{diagstates}
|{\Omega }_{\alpha }^{\Theta }\rangle =\sum_{j=1}^{m}C_{j\alpha }^{\Theta
}|\varphi _{j}^{n_{j}}\rangle
\end{equation}
which may account for additional correlations. Nevertheless, because many of
these correlations have already been accounted for in the MR expansion of
each of the $m$ basis states (as discussed above), one may expect the role
of this final diagonalization to be, in general, less important than in the
scheme used in Ref. 36.

Both the GHF-FED and GHF-EXC-FED VAP  approximations could be extended to the
use of general Hartree-Fock-Bogoliubov (HFB) transformations. \cite%
{Carlo-review} This, however, would require an additional particle number
symmetry restoration that increases our numerical effort by around one order
of magnitude and has hence not been included in the present study.

\subsection{Spectral functions and density of states}

\label{SFplusDOS} In this section, we briefly discuss the computation of the
SFs and DOS within our theoretical framework. Let us assume that for an $%
N_{e}$-electron system we have already obtained, along the lines described
in Sec.\ref{Symmetry-multiref-gs}, a GHF-FED ground state 
solution $|\phi_{1K}^{n_{1}\Theta }\rangle$. For all the
lattices considered in the present study the ground state has spin $S=0$ but
not necessarily linear momentum zero [i.e., $\Theta =(0,k)$]. In all cases, the ground
state transforms as an irrep of dimension 1.
Therefore, for this specific case, we can simply write the ground state wave
function as $|n_{1},N_{e},k\rangle $. The ground state energy will be
denoted as $E^{n_{1}k}$.

%
%
\begin{figure}[tbp]
\includegraphics[width=0.39\textwidth]{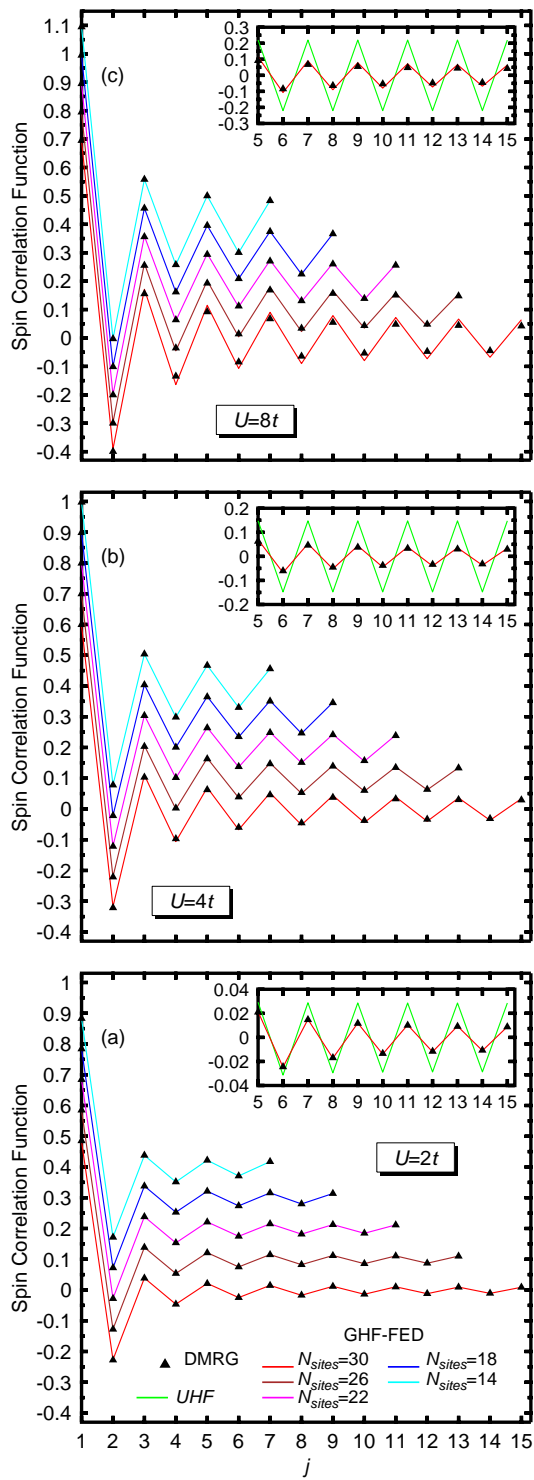}
\caption{(Color online) GHF-FED ground state spin-spin correlation functions in real
space  for half-filled lattices of different sizes (red, brown, magenta, blue and cyan curves).
DMRG values are plotted with black triangles.
Results are shown for U=$2t$ (a), $4t$ (b) and $8t$ (c). In each panel,
the inset displays a close-up of the long-range behavior predicted by the
GHF-FED (red curve) and DMRG (black triangles) schemes compared with the one obtained within the
standard UHF approximation (green curve) in the case of the $N_{sites}$=30
lattice. For more details, see the main text. }
\label{ss-N30-Ne30}
\end{figure}
%
%

Usually, the SFs are defined as the imaginary part of the time-ordered
Green's function and can be calculated from the Lehmann representation. \cite%
{Fetter-W} In order to compute them, we approximate \cite{Rayner-2D-Hubbard-PRB-2012,Carlos-Hubbard-1D} the ground states of 
the ($N_{e}\pm 1$)-electron systems with the quantum numbers ${\Theta }^{\pm}=(1/2,k^{\pm })$. For the $(N_{e}-1)$-electron system we superpose the
Goldstone (hole) manifolds $\hat{R}(\Omega )\hat{R}(g)\hat{b}_{h}({\mathcal{D}}_{1}^{i})|{\mathcal{D}}_{1}^{i}\rangle $ in the ansatz

\begin{equation} \label{hole-wf-spectral}
|n_{T},N_{e}-1,k^{-}\rangle =\sum_{ihM}f_{hM}^{i{\Theta }^{-}}\hat{P}_{KM}^{{%
\Theta }^{-}}\hat{b}_{h}({\mathcal{D}}_{1}^{i})|{\mathcal{D}}_{1}^{i}\rangle
\end{equation}
where $i=1,\dots ,n_{T}$. The number $n_{T}$ of
GHF-transformations used to expand the state Eq.(\ref{hole-wf-spectral}) may
be different from the one (i.e., $n_{1}$) in the GHF-FED ground state wave
function. We write $\hat{b}_{h}({\mathcal{D}}_{1}^{i})$ to explicitly
indicate that holes are made on different intrinsic determinants $|{\mathcal{D}}_{1}^{i}\rangle $ 
corresponding to the lowest-energy states of the $N_{e}$-electron system approximated by a single 
symmetry-projected
configuration along the lines
described in Sec.\ref{Hubbard1DHamiltonian}. The 
hole index $h$ runs as $h=1,\dots ,N_{e}$. Note, that the label $b=\pm 1$ is not
explicitly written in this section, but it is taken into account whenever
needed.

For the $(N_{e}+1)$-electron system we superpose the Goldstone (particle)
manifolds $\hat{R}(\Omega )\hat{R}(g)\hat{b}_{p}^{\dagger }({\mathcal{D}}_{1}^{i})|{\mathcal{D}}_{1}^{i}\rangle $ and write

\begin{equation} \label{particle-wf-spectral}
|n_{T},N_{e}+1,k^{+}\rangle =\sum_{ipM}g_{pM}^{i{\Theta }^{+}}\hat{P}_{KM}^{{%
\Theta }^{+}}\hat{b}_{p}^{\dagger }({\mathcal{D}}_{1}^{i})|{\mathcal{D}}%
_{1}^{i}\rangle
\end{equation}
where the index $i$ runs again as $i=1,\dots ,n_{T}$
and $p=N_{e}+1,\dots ,2N_{sites}$. The mixing coefficients $f^{i{\Theta }^{-}}$ and $g^{i{\Theta }^{+}}$ 
are determined by solving eigenvalue equations similar
to Eq.(\ref{HW-1}). This yields a maximun of $2n_{T}\times N_{e}\times d$
hole solutions with energies $E^{n_{T}k^{-}}$ and a maximum of $2n_{T}\times(2N_{sites}-N_{e})\times d$ particle solutions with energies $E^{n_{T}k^{+}}$
for each irreducible representation of the space group. The quantity $d$ is
the dimension of the corresponding irreducible representations, i.e., $d=1$
for $k^{\pm }=0,\pi $ and $d=2$ for $k^{\pm }\neq 0,\pi $. The hole ${\mathcal{B}}(q,\omega )$ and particle ${\mathcal{A}}(q,\omega )$ SFs are
then written in their standard form

\begin{align}  \label{SF-standard}
{\mathcal{B}}(q,\omega) =& \sum_{k^{-} \sigma} |\langle n_{T}, N_{e}-1,
k^{-} | {\hat{c}}_{q \sigma}| n_{1}, N_{e}, k \rangle |^{2}  \notag \\
\times& \delta \left(\omega - E^{n_{1} k} + E^{n_{T} k^{-}} \right)  \notag
\\
{\mathcal{A}}(q,\omega) =& \sum_{k^{+} \sigma} |\langle n_{T}, N_{e}+1,
k^{+} | {\hat{c}}_{q \sigma}^{\dagger} | n_{1}, N_{e}, k \rangle |^{2}
\notag \\
\times& \delta \left(\omega - E^{n_{T} k^{+}} + E^{n_{1} k}\right)
\end{align}
and the DOS can be computed as

\begin{eqnarray}  \label{EQ-DOS}
{\mathcal{N}}(\omega) = \sum_{q} \left({\mathcal{B}}(q,\omega) + {\mathcal{A}%
}(q,\omega) \right)
\end{eqnarray}

Due to the finite size of the considered lattices, both the hole and
particle SFs consist of a finite number of $\delta $ functions with
different weights. Therefore, we introduce an artificial Lorentzian width
$\Gamma$ for each state.

%
%
\begin{figure}[tbp]
\includegraphics[width=0.40\textwidth]{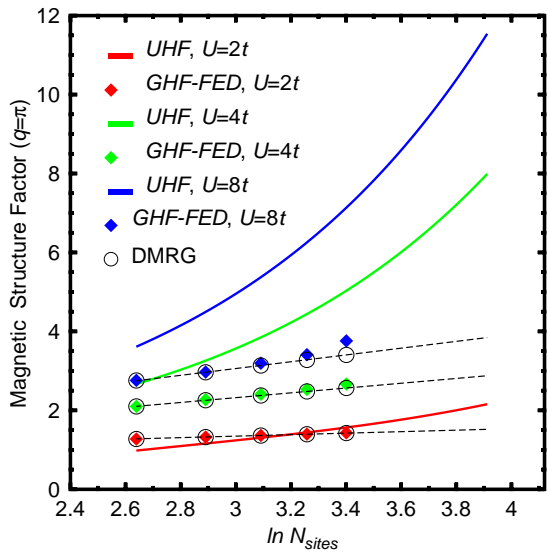}
\caption{(Color online) The GHF-FED magnetic structure
 factor, evaluated at the  
wave vector $q=\pi$, 
is plotted as a function of $ln$ $N_{sites}$ for 
half-filled lattices of 
different sizes. GHF-FED results are shown for on-site
repulsions of $U=2t$ (red diamonds), $4t$ (black diamonds),
and $8t$ (blue diamonds). The corresponding DMRG values are 
plotted with open circles. A straight line has been fitted to guide
the eye. The  magnetic structure factors predicted by the UHF
approximation for $U=2t$ (continuous red curve), $4t$ (continuous black
curve), and $8t$ (continuous blue curve) are also included for comparison
purposes.}
\label{SKK-vs-sites}
\end{figure}
%
%

\section{Discussion of results}

\label{Results}

In this section, we discuss the results of our calculations for some
illustrative examples. In most cases, we consider on-site repulsions of $U=2t,4t$ and $8t$ representing weak, intermediate-to-strong (i.e.,
non-interacting band width) and strong correlation regimes. First, in Sec.\ref{results-gs-energies}, we consider the ground states of half-filled
lattices of various sizes. We compare our ground
state and correlation energies with the exact ones, as
well as with those obtained using other theoretical approaches. We then
discuss the dependence of the predicted correlation energies on the number $%
n_{1}$ of non-orthogonal symmetry-projected configurations used to expand our ground state
wave functions. The computational performance of our scheme is also
addressed. The structure of the intrinsic GHF-determinants resulting from
our GHF-FED VAP  procedure is discussed in Sec.\ref{solitons-GHF-FED}. Our
results for SSCFs in real space and MSFs, for lattices with up to 30 sites, are
 presented in Sec.\ref{spinCFand MSF}. They are compared with DMRG 
results. For all the considered lattices, we have retained 1024 states 
in the renormalization procedure.
In Sec.\ref{DOSandSF}, we compare the DOS provided by our theoretical
framework with the exact one, obtained with an in-house full diagonalization
code in a lattice with $N_{sites}=10$. Hole SFs are also discussed in the
case of $N_{sites}=30$. Finally, in Sec.\ref{spectral}, we present results
obtained for the excitation spectra in lattices with $N_{sites}=12,14$ 
and $20$ and also discuss the structure of the underlying symmetry-broken
GHF-determinants resulting from our GHF-EXC-FED VAP procedure for excited
states.

\subsection{Ground state and correlation energies}

\label{results-gs-energies}

Let us start by considering lattices of 12 and 20 sites with the GHF-FED scheme
discussed in Sec.\ref{Symmetry-multiref-gs}. The corresponding $\Theta
=(0,\pi )$ ground states have $B_{1}$ symmetry, i.e., they are symmetric
under the reflection $x\rightarrow -x$. In Table I, we compare the predicted
ground state energies with
the exact ones.  For completeness, we also include energies
provided by the standard (i.e., one transformation) restricted (RHF) and
unrestricted (UHF) HF frameworks. Ours is a VAP approach
whose quality can be checked by studying how well it reproduces the exact
ground state correlation energies. To this end, we consider the ratio

\begin{equation} \label{formulaCE}
\kappa _{GHF-FED}=\frac{E_{RHF}-E_{GHF-FED}}{E_{RHF}-E_{EXACT}}\times 100
\end{equation}
between the GHF-FED and the exact correlation energies. For the UHF
approximation, $\kappa _{UHF}$ is obtained from a similar expression.

We observe from Table I that the inclusion of $n_{1}$=10 non-orthogonal
symmetry-projected configurations with the GHF-FED approach significantly improves
correlation energies with respect to UHF. In fact, $\kappa _{GHF-FED}\geq
99.79\%$ in all considered correlation regimes even for $N_{sites}=20$ which
is out of reach with exact diagonalization.

In the case of $N_{sites}=14$, whose $\Theta =(0,0)$ ground state has $A_{1}$
symmetry, i.e., it is symmetric under the reflection $x\rightarrow -x$, our
calculations with $n_{1}=10$ transformations predict 
energies of $-11.9539t$, $-8.0874t$, and $-4.6127t$ compared to 
the exact ones of $-11.9543t$, $-8.0883t$, and $-4.6131t$ for $U=2t,4t,$ and $8t$, respectively. This yields
$\kappa _{GHF-FED}$ values of $99.95,99.97$, and $99.99\%$, respectively.
Results for this lattice have been reported in the literature with the ResHF
framework using the half-projection method. \cite{Ikawa-1993} For
half-filled lattices with sizes comparable to the ones already mentioned,
the Gutzwiller method \cite{Gutzwiller_method} provides $\kappa $ ratios
around $85,77,$ and $50\%$, respectively.\cite{Gutzwiller1,VMC-Yokoyama} Let us also mention that our GHF-FED energies for
$N_{sites}=12$ and $14$ improve upon previously reported VAP values of $-6.9093t$ and $-8.0577t$ 
for $U=4t$. \cite{Carlos-Hubbard-1D} 

Calculations have also been carried out for $N_{sites}=16$, whose $\Theta
=(0,\pi )$ ground state has $B_{1}$ symmetry. We have obtained ground state
energies of $-16.4754t$ for $U=t$ and $-9.2122t$ for $U=4t$ while the exact
ones  are $-16.4758t$ and $-9.2144t$, respectively. Previous DMRG results 
for this lattice, have been reported in the literature. \cite{Xiang}
For all the lattices with sizes $N_{sites}$ $\le$ 18  
our DMRG calculations, retaining 1024 states in the renormalization 
procedure, reproduce the exact Lieb-Wu ground state energies
(to all quoted figures)
 for 
the considered U values. 
 
%
%
\begin{figure*}[tbp]
\includegraphics[width=1.\textwidth]{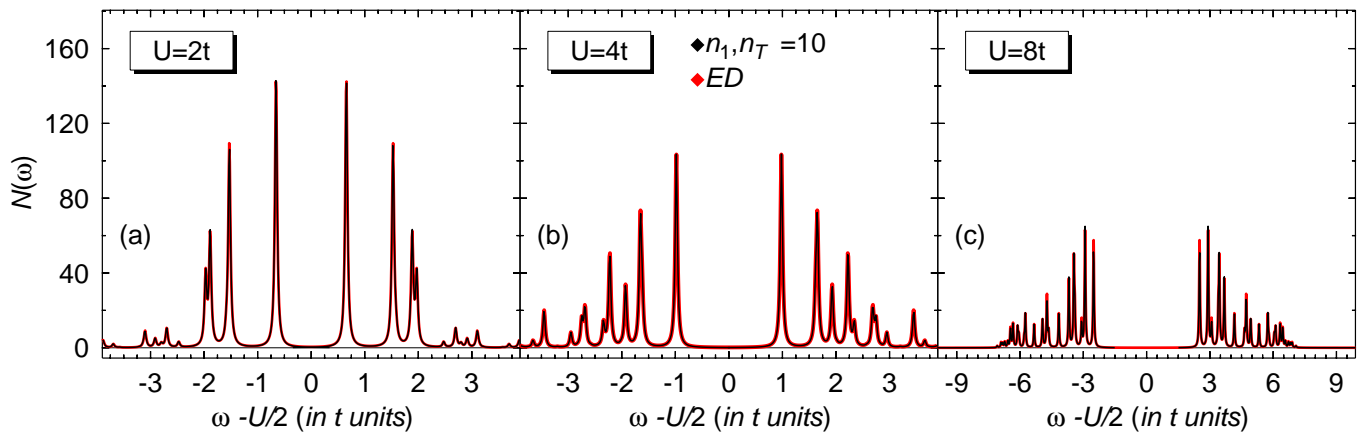}
\caption{(Color online) The DOS (black) for the half-filled lattice with 
$N_{sites}=10$ at $U=2t,4t,$ and $8t$ is plotted in panels (a), (b), and (c),
respectively, as a function of the shifted excitation energy $\omega-U/2$ (in $t$ units). Results have
 been obtained by approximating the $N_{e}$
and $(N_{e}\pm 1$)-systems
with $n_{1}=10$ and $n_{T}=10$ GHF-determinants. Our results are hardly
distinguishable from the DOS obtained with exact diagonalization (red). A
Lorentzian folding of width $\Gamma =0.05t$ has been used. For more details,
see the main text. }
\label{DOSN10Ne10}
\end{figure*}
%
%

The ground state energies 
for the lattices with $N_{sites}=30$ and $50$ are compared in Table II with the exact ones. In 
this case, the corresponding $\Theta =(0,0)$ ground
states have $A_{1}$ symmetry. In the same table, we also present ground
state energies predicted with the ResHF method \cite{Tomita-1} based 
on $n_{1}=30$ UHF-transformations (i.e., UHF-ResHF). It is very satisfying to
observe that both the GHF-FED and the UHF-ResHF VAP
 schemes can account 
for $\kappa \geq 98\%$ in a relatively large lattice with $N_{sites}=30$. In 
fact, the GHF-FED scheme provides $\kappa \geq
99.39\% $ with 45,048 variational parameters that represents a small
fraction of the dimension of the restricted (i.e., accounting for all
symmetries) Hilbert space in this lattice. In this case, our GHF-FED energy
also improves the variational value $-16.6060t$ obtained in Ref. 35 for $U=4t$ using 
a single symmetry-projected configuration. 
Note, that the ResHF method \cite{Fukutome-original-RSHF,Tomita-1} is not intrinsically limited to the use of
UHF-transformations and, therefore, the UHF-ResHF ground state energies
shown in Table II can still be improved by, for example, adopting
GHF-transformations as basic building blocks. \cite{Tomita-2011PRB} 
On the other hand, for $N_{sites}=30$ our DMRG calculations provide the energies
$-25.3830t$, $-17.2334t$, and $-9.8387t$ for $U=2t,4t,$ and $8t$, respectively.

Let us now comment on our results for $N_{sites}=32$ whose $\Theta =(0,\pi )$
ground state has $B_{1}$ symmetry. We have used $n_{1}=25$
GHF-transformations. For on-site
repulsions of $U=t$ and $2t$, we have obtained energies of $-33.2137t$ and
 $-26.9814t$ while the UHF-ResHF
  ones \cite{Tomita-1} are $-33.2128t$ and $-26.9556t$, respectively. These energies, should be compared with the exact
ones of $-33.2152t$ and $-27.0183t$ as well as with the
DMRG values  $-33.2141t$ and $-27.0177$. For previous DMRG calculations 
for this lattice the reader is referred to Ref. 62.

From the previous results, we conclude that the GHF-FED approximation can be
considered a reasonable starting point for building correlated ground state
wave functions which, at the same time, respect the original symmetries of
the 1D Hubbard Hamiltonian. This is further corroborated from the results,
shown in Table II, for $N_{sites}=50$. In particular, even when our
description of the ground state in this lattice is poorer than in the $%
N_{sites}=30$ case since we have kept the same number  of
GHF-transformations, it is remarkable that we obtain (with 125,048
variational parameters) the values $\kappa _{GHF-FED}=96.46,95.19,$ and $98.65\%$, respectively. For the same on-site
repulsions the variational Monte Carlo method \cite{VMC-Yokoyama} predicts $\kappa $ values of
around $87,92,$ and $96\%$. The corresponding UHF-ResHF values \cite{Tomita-1} are also listed in Table II.

%
%
\begin{figure*}[tbp]
\includegraphics[width=1.0\textwidth]{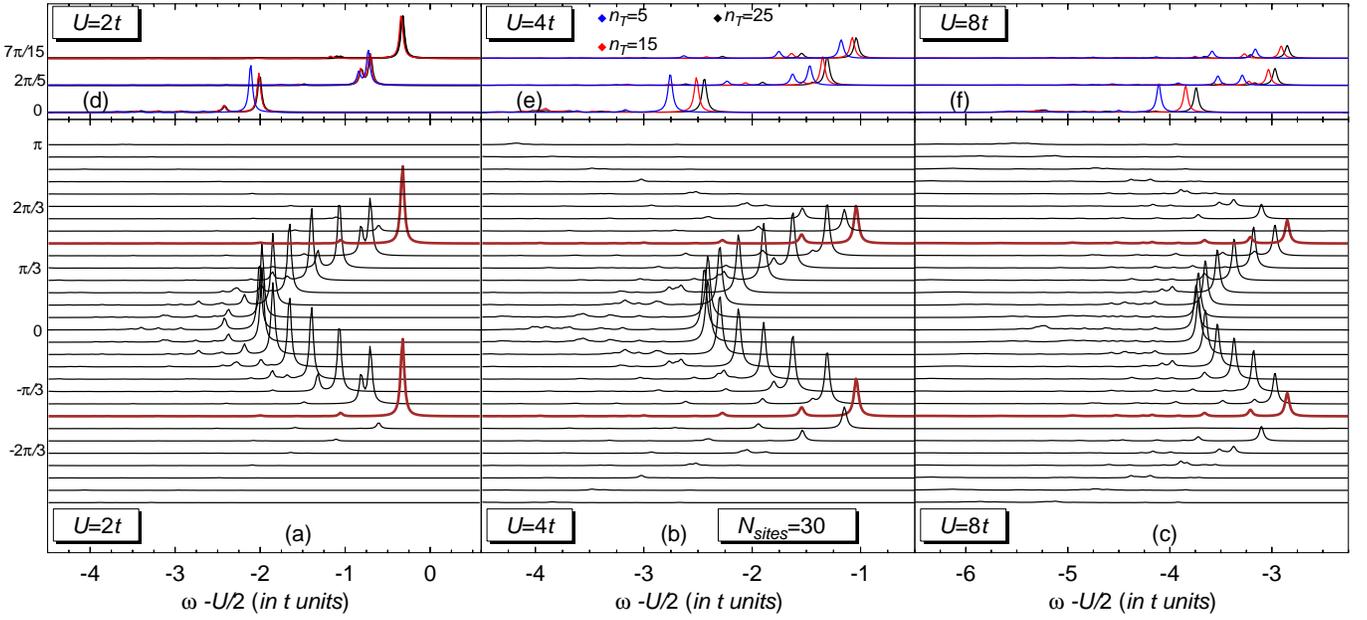}
\caption{(Color online) The hole SFs 
for the half-filled lattice with $N_{sites}=30$ at $U=2t,4t,$ and $8t$ are plotted in panels (a), (b),
and (c) as functions of the shifted excitation energy $\omega -U/2$
(in $t$ units). Results have been obtained by approximating the $N_{e}$ and 
$(N_{e}\pm 1$)-systems with $n_{1}=25$ and $n_{T}=25$
GHF-determinants. The hole SFs for momenta identical to the Fermi momentum
are displayed in brown color. The shapes of some selected hole 
SFs (i.e., $k=0, 2\pi /5$ and $7\pi /15$), obtained by approximating the ground 
states of the $(N_{e}\pm 1$)-systems
with $n_{T}=5$
(blue), $15$ (red) and $25$ (black) 
but the ground state of the $N_{e}$-system always with $n_{1}$=25
GHF-transformations, are compared in
panels (d), (e), and (f). A Lorentzian folding of width $\Gamma =0.05t$ has
been used. For more details, see the main text.}
\label{SFholesN30}
\end{figure*}
%
%

In Fig.\ref{energy_vs_trans}, we have plotted the ratio $\kappa _{GHF-FED}$, as a function of the inverse $1/n_{1}$ of the number
of transformations $n_{1}$ included in the GHF-FED ansatz, for lattices with $N_{sites}=20$ 
and $30$. They
increase smoothly with the number of non-orthogonal 
symmetry-projected
configurations used
to expand the wave function. From Fig.\ref{energy_vs_trans}, it 
is apparent (see also Tables I and II) that with
increasing lattice size, we need a larger number $n_{1}$ of
symmetry-projected configurations to keep and/or improve the quality of the GHF-FED wave
functions. For example, comparing the $N_{sites}=20$ and $30$ lattices, we
see that in the former $n_{1}=10$ transformations are enough to obtain $\kappa _{GHF-FED}\geq 99.79\%$ 
while in the latter $98.69\%\leq \kappa _{GHF-FED}\leq 99.49\%$. On the other hand, in 
the $N_{sites}=50$ case, $n_{1}=10$ transformations leads to $93.37\%\leq \kappa_{GHF-FED}\leq 97.68\%,$ whereas with 
$n_{1}=25$ we reach the $\kappa_{GHF-FED}$ values shown in Table II.

Obviously, as in many other approaches to many-fermion systems, we are
always limited to a finite number of configurations in practical
calculations. Nevertheless, the GHF-FED scheme provides compact ground state
wave functions whose quality can be systematically improved by adding new
(variationally determined) non-orthogonal symmetry-projected configurations. In fact, both
ours and the ResHF \cite{Yamamoto-2,Ikawa-1993,Tomita-1} wave functions are
nothing else than a discretized form of the exact coherent-state
representation of a fermion state \cite{Perlemov} and, therefore, become
exact in the limit $n_{1}\rightarrow \infty $. Our aim in the present work
is not to lower the ground state energy as much as possible but to test to
which extent our scheme can account for relevant correlations in the
considered lattices. Therefore, for the largest lattices here studied (i.e.,
$N_{sites}=30$ and $50$), we have restricted ourselves in practice to a
maximun number $n_{1}=25$ of GHF-transformations.

A few words concerning the computational performance of our method are in
order here. In panel (a) of Fig.\ref{sppedupplot}, we have plotted the
speedup of a typical calculation as a function of the number of proccessors.
Results are shown for $N_{sites}=50$ and $U=4t$ but similar behavior was
also obtained for $U=2t$ and $8t$. As demonstrated in the plot, the GHF-FED
speedup grows linearly with the number of processors used in the
calculations. On the other hand, panel (b) of the same figure shows that
(for a fixed number of proccessors) an efficient implementation of our
variational scheme scales linearly with the number $n_{1}$ of
GHF-transformations used 
while the ResHF scaling is quadratic. Concerning the scaling of our 
method with system size, Fig.\ref{energy_vs_trans} shows that as the system becomes 
larger a larger number of transformations is required to keep 
 the quality of our wave functions. We cannot currently determine 
 how the number of transformations scales 
 with system size as this would require to consider 
 larger lattices than the ones studied in this paper.

\subsection{Structure of the intrinsic determinants and basic units of
quantum fluctuations}

\label{solitons-GHF-FED}

An interesting issue is whether there is any relevant information in
the symmetry-broken (i.e., intrinsic) determinants
$|{\mathcal{D}}_{1}^{i}\rangle $ resulting from the GHF-FED VAP
optimization. We are interested in comparing the structure of these
determinants with the spin-density wave solution obtained with the
standard UHF approximation.  Here, one should keep in mind that a
variationally optimized GHF-determinant has the same energy as the
optimal UHF one. \cite{Bach-Lieb-Solovej} We have studied the quantity

\begin{equation} \label{SDW1}
{\xi }_{1}^{i}(j)=(-)^{j-1}\langle {\mathcal{D}}_{1}^{i}|\mathbf{S}(j)|{%
\mathcal{D}}_{1}^{i}\rangle \cdot \langle {\mathcal{D}}_{1}^{i}|\mathbf{S}%
(1)|{\mathcal{D}}_{1}^{i}\rangle
\end{equation}
where $j=1,\dots ,N_{sites}$ is the lattice index while $i=1,\dots ,n_{1}$
enumerates the GHF-determinants in the GHF-FED ground state (subscript $1$) solution. Among 
the $n_{1}=25$ transformations ${\mathcal{D}}_{1}^{i}$ used 
for the $N_{sites}$=50 lattice, we have selected
some typical examples to plot the quantity ${\xi }_{1}^{i}(j)$. Results are
displayed in panels (a) to (d) of Fig.\ref{solitonsU2}. Other
determinants $|{\mathcal{D}}_{1}^{i}\rangle $, not shown in the figure,
exhibit the same qualitative features. Similar results are also found for other 
on-site repulsions, as well as for other lattices.

%
%
\begin{figure}[tbp]
\includegraphics[width=0.45\textwidth]{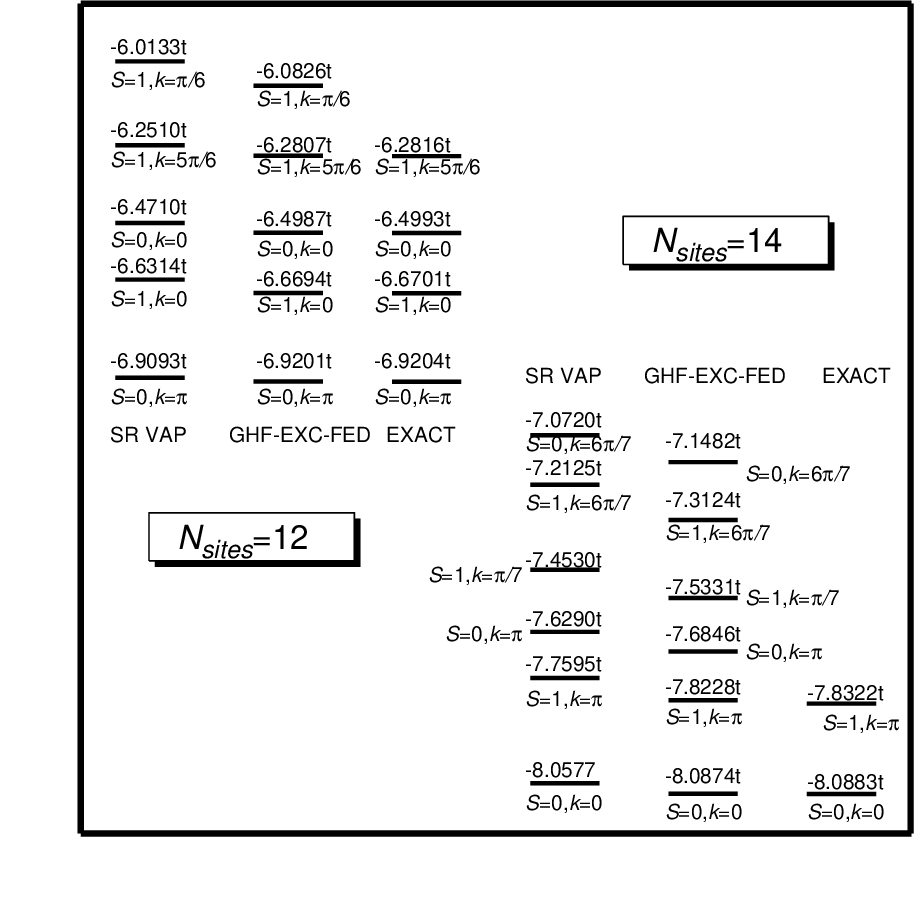}
\caption{The energies of some selected states obtained within the
GHF-EXC-FED approximation for the half-filled lattices with $N_{sites}=12$ and $14$ are compared with the
ones provided by single-reference VAP
(SR VAP) calculations (only 3D spin and linear momentum
projections) as well as with the exact ones from Lanczos diagonalization. \cite{Carlos-Hubbard-1D} 
Results are shown for $U=4t$. }
\label{VAMPIR_vs_FED}
\end{figure}

%
%

For the standard UHF spin-density wave solution, the quantity ${\xi }_{1}^{i}(j)$ has
nearly constant positive values plotted with red lines in Fig.\ref%
{solitonsU2}. A very different behavior appears in the intrinsic
GHF-determinants $|{\mathcal{D}}_{1}^{i}\rangle $ associated with the
GHF-FED solution. First, we observe a broad
spin feature distributed all over the lattice, which is a consequence of the
richer spin textures provided by the use of GHF-transformations and full 3D
spin projection. In addition, pairs of points (black squares) appear where ${\xi }_{1}^{i}(j)$ changes 
its sign (i.e., the spin-density wave reverses it phase). These
defects of the spin-density wave phase represent soliton-antisoliton ($S-\overline{S}$)
pairs in the case of half-filled lattices. \cite{Yamamoto-1,Ikawa-1993,Tomita-1,Horovitz} In particular, our analysis of the
charge densities $\rho _{1}^{i}(j)=1-\sum_{\sigma }\langle {\mathcal{D}}_{1}^{i}|\hat{n}_{j\sigma }|{\mathcal{D}}_{1}^{i}\rangle $ reveal that they
correspond to neutral $S^{0}-\overline{S^{0}}$ pairs. Let us stress that the
presence of at least one $S^{0}-\overline{S^{0}}$ pair is a genuine VAP
effect appearing even if we approximate a given ground state within a SR
framework, \cite{non-unitary-paper-Carlos} as discussed in Sec.\ref{Hubbard1DHamiltonian}.

Furthermore, Fig.\ref{solitonsU2} illustrates how the $S^{0}-\overline{S^{0}}$ pairs appear
 at different lattice locations $j$ with varying distance $R_{S^{0}-\overline{S^{0}}}$ among the members of the pairs. 
The latter represents the breathing
motion of the $S^{0}-\overline{S^{0}}$ pairs. The $S^{0}-\overline{S^{0}}$
pairs are present in all the intrinsic determinants $|{\mathcal{D}}_{1}^{i}\rangle $ associated 
with the GHF-FED expansion, which as already mentioned above, superposes the
Goldstone manifolds $\hat{R}_{S}(\Omega )\hat{R}(g)|{\mathcal{D}}_{1}^{i}\rangle $ containing 
defects in the spin-density wave. We are then left with an
intuitive physical picture in which the
soliton pairs
can be regarded as basic units of quantum fluctuations in our GHF-FED
states. On the other hand, the interference between
$S^{0}-\overline{S^{0}}$ pairs belonging to different symmetry-broken
determinants $|{\mathcal{D}}_{1}^{i}\rangle $ is accounted for in our
calculations through a resonon-like equation similar to
Eq.(\ref{HW-1}). This interpretation has been suggested in previous
studies with the ResHF method.
\cite{Yamamoto-1,Yamamoto-2,Ikawa-1993,Tomita-1}

\subsection{Spin-spin correlation functions and magnetic structure factors}

\label{spinCFand MSF}

Let us now consider the ground state spin-spin correlation functions (SSCFs)
in real space. For a given set of symmetry quantum numbers $\Theta $, they
can be computed  as

\begin{eqnarray}  \label{ss-CF}
{\mathcal{F}}_{m}^{n_{1} \Theta}(j) = \frac{ \langle \phi_{1 K}^{n_{1}
\Theta} | \mathbf{S}(j) \cdot \mathbf{S}(1) | \phi_{1 K}^{n_{1} \Theta}
\rangle } { \langle \phi_{1 K}^{n_{1} \Theta} | \phi_{1 K}^{n_{1} \Theta}
\rangle }
\end{eqnarray}

Note that if a wave function has good spin, as it is the case 
with the GHF-FED one, the
SSCFs have to be the same for all the members of a (2S+1)-multiplet and,
therefore, they cannot depend on the $\Sigma $ quantum 
number.
However, a dependence with respect to the particular row $m$ of the space
group irreducible representation that we are using in the projection still remains and is 
explicitly included in ${\mathcal{F}}_{m}^{n_{1} \Theta}(j)$.

The SSCFs corresponding to the ground states for $N_{sites}=14,18,22,26,$
and $30$, approximated by $n_{1}=10,10,15,25,$ and $25$ GHF-transformations,
respectively, are depicted in panels (a), (b) and (c) of Fig.\ref{ss-N30-Ne30}. In 
the same figure, we have also plotted
the values resulting from our DMRG calculations. We observe a  good agreement 
between the GHF-FED and DMRG SSCFs with slight deviations for the 
$N_{sites}=30$ lattice at U=8t, which can be improved 
by increasing the number of transformations. In particular, both SSCFs display a rapid decrease
for $j\leq 3$. A similar feature has been studied 
in previous works. \cite{Tomita-1,Yamamoto-1,Yamamoto-2,Ikawa-1993,Shiba-Pincus} Regardless of the
on-site interaction, the short range part of the SSCFs runs parallel for the
lattices considered in Fig.\ref{ss-N30-Ne30}, pointing to converged behavior
as a function of lattice size. Morevover, the mid and long range amplitude
of the SSCF for a given lattice increases with increasing $U$ values.

In each panel of Fig.\ref{ss-N30-Ne30}, the inset displays a close-up of the
long range behavior of the SSCF predicted by the GHF-FED
and DMRG 
 approximations 
compared with the one obtained within the standard UHF approach
for $N_{sites}=30$. As can be observed, the amplitude of the
UHF SSCF remains constant for $j\geq 5$ while the GHF-FED and DMRG ones exhibit a
damped long range trend. Previous studies have suggested that there are two important
ingredients necessary to account for a qualitatively correct long range
behavior of the SSCFs: the self-consistent optimization of the intrinsic
determinants (i.e., orbital relaxation \cite{Yamamoto-1,Ikawa-1993}) and
having pure spin states (i.e., no spin contamination \cite{Tomita-1}). Our
wave functions meet both conditions.

%
%
\begin{figure*}[tbp]
\includegraphics[width=1.\textwidth]{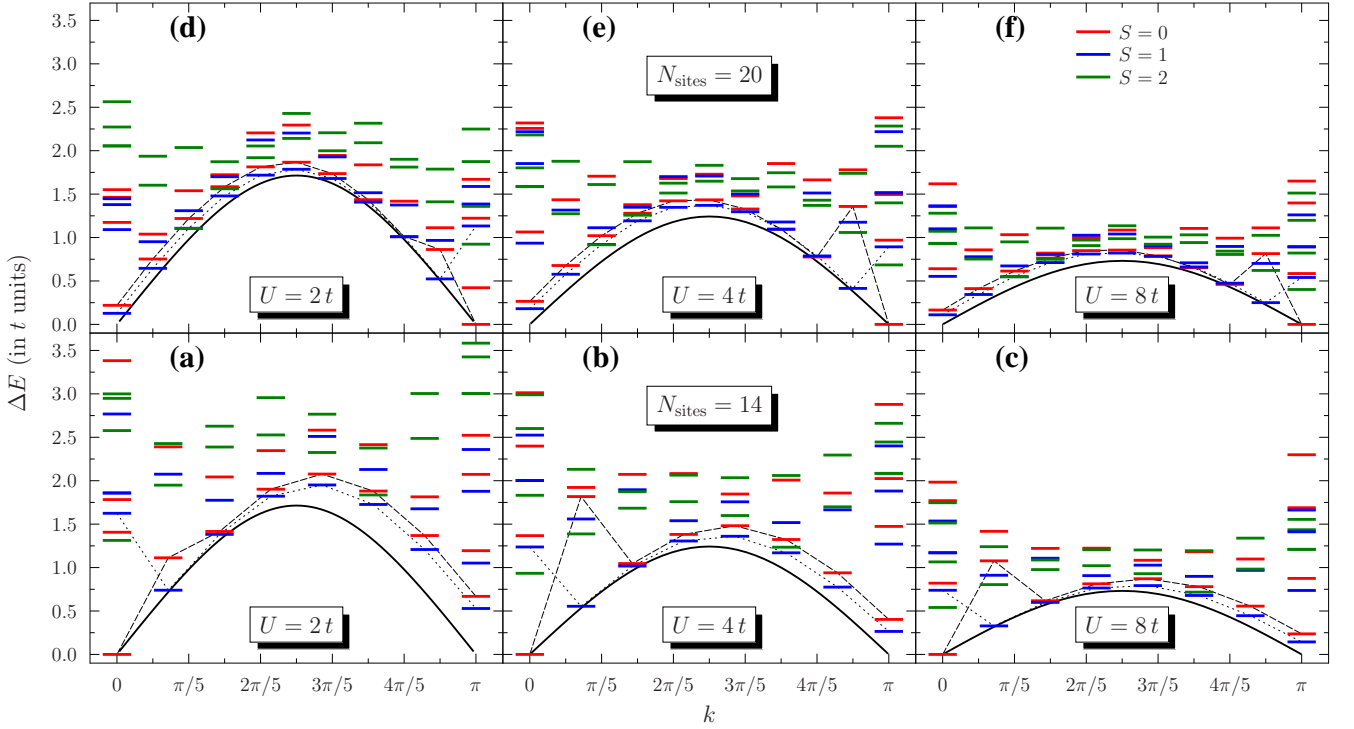}
\caption{(Color online) Energy spectrum obtained via Eq.(\protect\ref%
{diagonalizationfinal}) for the half-filled lattices with $N_{sites}=14$ [panels (a), (b), and (c)] and $20
$ [panels (d), (e), and (f)]. For each irreducible representation of the
space group, the lowest-energy and first excited states with the spins $S=0$
(red bars),$1$ (blue bars), and $2$ (green bars) have been plotted. Results
are shown for $U=2t,4t,$ and $8t$. The exact dispersion curves \protect\cite%
{exactkdisp} for $N_{sites}\rightarrow \infty $ (thin black lines) are also
included. In order to guide the eye, the lowest-lying states with spin $S=0$
($S=1$) have been connected by long (short) dashed lines. For more details,
see the main text. }
\label{FED-dispersions}
\end{figure*}

%
%

The magnetic structure factors (MSFs), evaluated at the  
wave vector $q=\pi$, can be computed as

\begin{equation} \label{magnetic-SKK}
S_{m}^{n_{1}\Theta }(\pi)=\frac{1}{N_{sites}}\sum_{ij}(-)^{i+j}\frac{\langle
\phi _{1K}^{n_{1}\Theta }|\mathbf{S}(i)\cdot \mathbf{S}(j)|\phi
_{1K}^{n_{1}\Theta }\rangle }{\langle \phi _{1K}^{n_{1}\Theta }|\phi
_{1K}^{n_{1}\Theta }\rangle }
\end{equation}
and the ones corresponding to the ground states for $N_{sites}=14,18,22,26,$
and $30$ are displayed in Fig.\ref{SKK-vs-sites} as functions of $ln$ $%
N_{sites}$. The corresponding DMRG results are shown in the same plot. 
We have also included the UHF MSFs for comparison purposes. At
variance with the UHF MSFs which diverge exponentially, both the  GHF-FED
and DMRG
 results
display an almost linear behavior. A previous 
work\cite{Imada1overrMSF} has
shown that the SSCFs in real space behave for a half-filled system as $%
\approx (ln^{\sigma }j)/j$. This implies that as a funcion of the lattice
size, the MSFs should behave as $ln^{1+\sigma }$ $N_{sites}$. In
Fig.\ref{SKK-vs-sites}, we have simply fitted a straight line using the DMRG MSFs 
to guide the eye. We
have not attempted to determine logarithmic corrections as this would
require larger lattices than those studied in the present paper.

\subsection{Spectral functions and density of states}
\label{DOSandSF} 

In panels (a), (b), and (c) of Fig.\ref{DOSN10Ne10}, we
have plotted (black) the DOS ${\mathcal{N}}(\omega )$  for $N_{sites}=10$. 
In the same figure, we have also plotted (red) the exact DOS
obtained with an in-house full diagonalization code. There is excellent
agreement between ours and the exact  DOS concerning the position and relative
heights of all the prominent peaks. Both ours and the exact DOS exhibit the
particle-hole symmetry well known for half-filled systems \cite{text-Hubbard-1D} 
and a splitting into the lower and upper Hubbard bands.
The Hubbard gap between these bands increases with larger $U$. Both
dynamical cluster approximation \cite{moukouri2001,huscroft2001,aryanpour2003}  and cellular dynamical mean
field theory \cite{AraGo}  studies suggest that this gap is preserved
for any finite value of the on-site interaction at sufficiently low
temperatures even in the thermodynamic limit, with $U=0t$ being the
only singular point.

Tendencies to spin-charge separation as well as other relevant shadow
features inside the Brillouin zone, similar to the ones expected in the
infinite-$U$ limit of the 1D Hubbard model, \cite{PENcPRL1996} have been
found in previous cellular dynamical mean
field theory \cite{AraGo} and cluster perturbation theory \cite{Senechal-scs} studies of the spectral weigths in the case of finite
on-site repulsions. In panels (a), (b), and (c) of Fig.\ref{SFholesN30}, we
have plotted the hole SFs for the $N_{sites}=30$ lattice.

The first feature observed from Fig.\ref{SFholesN30} is the Hubbard gap
opening at the Fermi momentum $k_{F}=7\pi /15$. The spectral weight
concentrates on the prominent peaks belonging to the spinon band. Our
calculations for smaller lattices with $N_{sites}=14$ and $20$ indicate that
this spinon band is quite stable in terms of lattice size although the
relative height of its peaks decreases for increasing $U$ values. The holon
singularities are clearly visible in some of the SFs shown in Fig.\ref%
{SFholesN30} for linear momenta $-\pi /2<k<\pi /2$. On the other hand, the
holon bands can also be followed for linear momenta $k>\pi /2$ and $k<-\pi /2
$. They are the mirror images of the ones with opposite $\omega -U/2$ values
\cite{Senechal-scs} and become apparent for $U=4t$ and $8t$. However,
besides the spinon band, the most relevant feature in our SFs is the very
extended distribution of the spectral weight for linear momenta $-\pi /2<k<\pi /2$. The 
comparison with our SFs for $N_{sites}=14$ and $20,$
obtained with $n_{1}=10$ and $n_{T}=10$ GHF-transformations, reveals that
the increase of lattice size produces more pronounced shadow features due to
the fragmentation of the spectral strength over a wider interval of $\omega
-U/2$ values. The previous finite size results show that our SFs exhibit
tendencies beyond a simple quasiparticle distribution and agree
qualitatively well with the ones obtained using other approximations. \cite{AraGo,Senechal-scs}

%
%
\begin{figure*}[tbp]
\includegraphics[width=1.\textwidth]{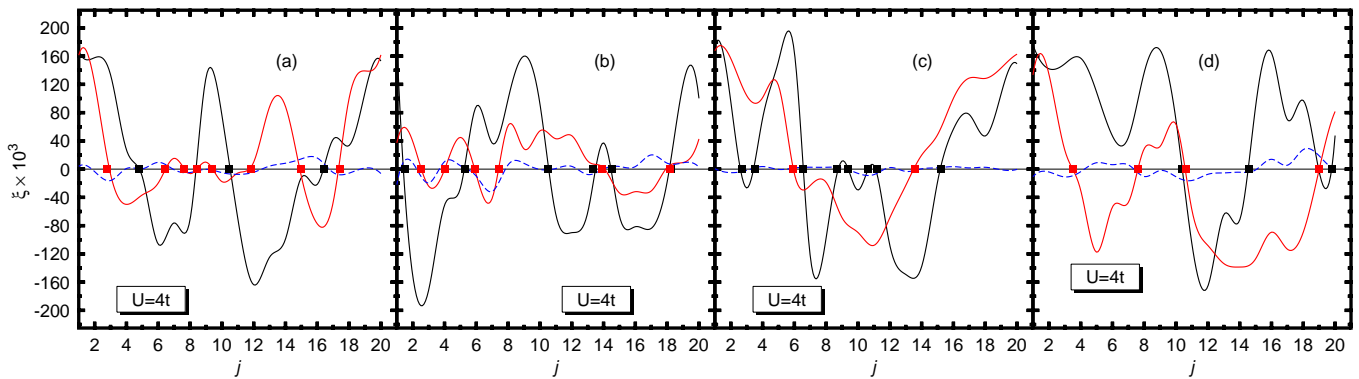}
\caption{(Color online) Structure of some typical symmetry-broken
GHF-determinants $|{\mathcal{D}}_{1}^{i}\rangle $ (black) and $|{\mathcal{D}}_{2}^{i}\rangle $ (red) 
used to expand the lowest-energy and first excited
states with spin $S=1$, linear momentum $k=0$ and $A_{1}$ symmetry. The
charge density (dashed blue) corresponding to the determinants $|{\mathcal{D}}_{2}^{i}\rangle $ is 
also included in each panel. Results are shown for
the half-filled lattice with $N_{sites}=20$ 
at
 $U=4t$. For more details, see the main text.}
\label{solitons_excited_states}
\end{figure*}

%
%

The shapes of some selected hole SFs are compared in panels (d), (e), and (f) of
 Fig.\ref{SFholesN30}. As 
can be seen from panel (d), $n_{T}=15$ transformations are
enough to account for all relevant details of the SFs shown in panel (a) for
$U=2t$. On the other hand, panels (e) and (f) show that a larger number of
transformations is required for $U=4t$ and $8t$. In particular, increasing
the number of transformations  for the ($N_{e}\pm 1$)-systems from $n_{T}=5$ to $n_{T}=15$ and/or $25$ leads to a shift of the main peaks and
redistributes the spectral strength of some of the peaks found in the SFs
for $n_{T}=5$ as a result of the small number of configurations used in the
calculations. This explains the differences between ours and the SFs
reported, for the same lattice at $U=4t$, in the previous 
VAP
 study \cite{Carlos-Hubbard-1D} of the 1D Hubbard model using only $n_{1}=1$ and $n_{T}=5
$ GHF-transformations.

\subsection{Excitation spectra}

\label{spectral} In this section, we consider the low-lying excitation
spectra obtained for $N_{sites}=12,14,$ and $20$ with the GHF-EXC-FED scheme
discussed in Sec.\ref{Symmetry-multiref-excited}. For each irreducible
representation of the space group, we have computed the lowest-energy and
first excited states with spins $S=0,1,$ and $2$. Each of these states has
been approximated by $10$ non-orthogonal 
symmetry-projected
 GHF-determinants. A final $%
2\times 2$ diagonalization  of the 1D Hubbard
Hamiltonian has also been carried out. For these particular lattices, we
have found that for each symmetry $\Theta $, the (Gram-Schmidt
orthonormalized) ground $|\varphi _{1K}^{n_{1}=10,\Theta }\rangle $ and
first excited $|\varphi _{2K}^{n_{2}=10,\Theta }\rangle $ states are very
weakly coupled through the Hamiltonian. Due to this, the energies
corresponding to the states $|{\Omega }_{1K}^{\Theta }\rangle $ and 
$|{\Omega }_{2K}^{\Theta }\rangle $ resulting from the $2\times 2$
diagonalization  are almost identical to
those corresponding to the basis states $|\varphi _{1K}^{n_{1}=10,\Theta}\rangle$
 and $|\varphi _{2K}^{n_{2}=10,\Theta }\rangle $. However, this
cannot be anticipated \textit{a priori} and the final 
diagonalization of the Hamiltonian 
should always be carried out.

In Fig.\ref{VAMPIR_vs_FED}, we compare the energies of some selected states
for $N_{sites}=12$ and $14$ with the ones obtained  in the
previous variational study \cite{Carlos-Hubbard-1D} of the 1D Hubbard 
model  where 3D
spin and linear momentum projections were carried out. The exact results in
Fig.\ref{VAMPIR_vs_FED} correspond to Lanczos diagonalizations. As can be
observed, our MR calculations, where in addition to 3D spin projection the
full space group of the 1D Hubbard model is taken into account, improve the
energies reported in Ref. 35 for both ground and excited states.

In Fig.\ref{FED-dispersions}, we show the low-lying spectrum obtained via
Eq.(\ref{diagonalizationfinal}), for $N_{sites}=14$ [panels (a), (b), and
(c)] and $20$ [panels (d), (e), and (f)] taken as representatives examples
of systems whose $\Theta =(0,0)$ and $\Theta =(0,\pi )$ ground states have $%
A_{1}$ and $B_{1}$ symmetries, respectively. 
We observe that both the lowest-lying singlet and triplet states obtained in
our calculations nicely follow the sine-like dispersion trend in the exact
curve for $N_{sites}\rightarrow \infty $. The anomaly observed in the
GHF-EXC-FED $k$-dispersion for the lowest-energy singlets and triplets has
also been found in previous studies within the ResHF framework \cite%
{Ikawa-1993} as well as in finite versions of the exact Lieb-Wu solutions.
\cite{Hashimoto} For any finite U value, the exact $N_{sites}\rightarrow
\infty $ curves exhibit gapped excitations, exception made for the $k=0$ and
$k=\pi $ states which are degenerate. In our calculations such a degeneracy
is broken due to finite size effects. However, we observe that for a given
finite $U$ value, the energy difference between the lowest singlet and
triplet states decreases with increasing lattice size. For example, for $U=2t
$ we have obtained $\Delta E_{s-t}=0.5287t$ in the case $%
N_{sites}=14$ while $\Delta E_{s-t}=0.1275t$ for $N_{sites}=20$%
. For increasing on-site respulsions, irrespective of the lattice size, an
overall compression of the spectra takes place. This is consistent with the
fact that in the limit $U\rightarrow \infty $ all the configurations shown
in Fig.\ref{FED-dispersions} should become degenerate.

\subsection{Structure of the intrinsic determinants and basic units of
quantum fluctuations in the GHF-EXC-FED wave functions}

\label{solitons-GHF-EXC-FED}

In Sec.\ref{solitons-GHF-FED}, we have discussed the structure of the
intrinsic determinants associated with the GHF-FED states. Here, we pay attention to the symmetry-broken ones
used to expand the GHF-EXC-FED wave functions. To illustrate our results, we consider
states belonging to the spectrum shown in panel (e) of Fig.\ref{FED-dispersions}. In particular, we have
plotted in panels (a) to (d) of Fig.\ref{solitons_excited_states} the
quantities ${\xi }_{1}^{i}(j)$ (black) and ${\xi }_{2}^{i}(j)$ (red)
computed [see, Eq.(\ref{SDW1})] with some of the $n_{1}=10$ and $n_{2}=10$
symmetry-broken determinants $|{\mathcal{D}}_{1}^{i}\rangle $ and $|{\mathcal{D}}_{2}^{i}\rangle $ used to expand 
the lowest-energy $|\varphi_{1K}^{n_{1}=10,\Theta }\rangle $ and first excited $|\varphi_{2K}^{n_{2}=10,\Theta }\rangle $ states
 with $\Theta =(1,0)$ and $A_{1}$
symmetry. Other determinants, not shown in the figure, exhibit the same
qualitative features. Similar results are also obtained for $U=2t$ and $8t$
as well as for other lattices.

As can be observed, both ${\xi }_{1}^{i}(j)$ and ${\xi }_{2}^{i}(j)$
display defects similar to the ones already discussed for the $S=0$
ground states provided by the GHF-FED approximation (see,
Fig.\ref{solitonsU2}). From this we conclude that not only the ground
but also the excited state wave functions provided by our MR VAP
scheme superpose Goldstone manifolds built in terms of intrinsic
GHF-determinants containing defects (i.e., solitons)
that can be regarded as basic units of
quantum fluctuations. In general, the intrinsic determinants
associated with different symmetry-projected states may develop local
variations of the charge density as seen (dashed blue curve) from
Fig.\ref{solitons_excited_states} where we have also plotted the
quantity $\rho _{2}^{i}(j)=1-\sum_{\sigma }\langle
{\mathcal{D}}_{2}^{i}|\hat{n}_{j\sigma }|{\mathcal{D}}_{2}^{i}\rangle
$.

\section{Conclusions}

The accurate description of the most relevant correlations in the ground and
low-lying excited states of a given many-fermion system, with as few
configurations as possible, is a central problem in quantum chemistry, solid
state, and nuclear structure physics. In the present study, we have explored
a VAP MR avenue for the 1D Hubbard model.
The main acomplishments of the present work are listed below.

(i) We have presented a powerful methodology of a VAP MR configuration
mixing scheme, originally devised for the nuclear many-body problem,
but not yet considered to study ground and excited states, with well
defined symmetry quantum numbers, of the 1D Hubbard model with
nearest-neighbor hopping and periodic boundary conditions. Both ground
and excited states are expanded in terms of non-orthogonal and
Ritz-variationally optimized symmetry-projected configurations. The
simple structure of our projected states allows an efficient
parallelization of our variational scheme, which scales linearly with
the number of processors as well as with the number of transformations
used in the calculations. The method also provides a (truncated) basis
consisting of a few Gram-Schmidt orthonormalized states. This basis
may be used to diagonalized the Hamiltonian to account, in a similar
fashion, for additional correlations in the ground and excited states
with well defined symmetry quantum numbers.

(ii) We have shown that our MR approximation gives accurate ground state
energies and correlation energies as compared with the exact Lieb-Wu
solutions  for relatively large half-filled lattices up to $%
30$ and $50$ sites. The comparison with other theoretical approaches also
reveals that our scheme can be considered as a reasonable starting point
for obtaining correlated ground state wave functions in the case of the 1D
Hubbard model. We have computed the full low-lying spectrum for 
the $N_{sites}=14$ and $20$ lattices. The momentum dispersion of the lowest-lying
singlet and triplet states follows the exact shape predicted by the Lieb-Wu
solution  in the thermodynamic limit. With increasing $U$
we also observe a general compresion of the spectrum.

iii) From the analysis of the structure of the intrinsic determinants
associated with our MR ground and excited state wave functions, we observe
that they all contain defects (i.e., solitons)
that can be regarded as basic units of quantum fluctuations for
the considered lattices.

(iv) Our results for the ground state SSCFs in real space show long range
decay that is not observed in the UHF case. The MSFs computed from such
correlation functions show a behavior approximately linear in $ln$ $N_{sites}
$ consistent with previous results available in the literature.

(v) Our approximation also allows to compute SFs and the DOS. To this
end, we considered ans\"{a}tze, whose flexibility is determined by the
numbers $n_{1}$ and $n_{T}$ of HF-transformations used to expand the
wave functions of systems with $N_{e}$ and $(N_{e}\pm
1)$-electrons. For a small lattice with $N_{sites}=10$ we have
compared the DOS predicted within our approach with the one obtained
using a full diagonalization and found an excellent agreement between
both. For a larger lattice with $N_{sites}=30$ our scheme provides
hole SFs that agree qualitatively well with the ones obtained with
other approximations and exhibit tendencies beyond a simple
quasiparticle distribution.

We believe that the finite size calculations discussed in the present study
already show that 
VAP
 approximations, based on MR expansions in terms of
non-orthogonal symmetry-projected HF-determinants, represent useful tools that complement
other existing approaches to study the physics of low-dimensional correlated
electronic systems. Within this context, the scheme presented in this work
leaves ample space for further improvements and research. First, the number
of non-orthogonal symmetry-projected configurations used in the corresponding MR expansions
can be increased to improve the quality of our wave functions. Second, we
could still incorporate particle number symmetry breaking (i.e., general
HFB-transformations) and restoration (i.e., particle number projection) to
access even more correlations. Third, our scheme can be easily extended to
the 2D case as well as to doped systems with arbitrary on-site interaction
strengths. Our approximation is also general enough so as to be implemented
for the molecular Hamiltonian \cite{CRG-QC} as well as for lattices like the
honeycomb, the Kagome or the Shastry-Sutherland \cite{S-S-LAttice} ones.
The same VAP MR scheme can also be applied to study frustrated Hubbard
models in the 1D and 2D cases. Finally, the MR scheme
discussed in the present work could also be used as a powerful solver in the
framework of fragment-bath embedding approximations. \cite{Knizia-Chan}
In particular, it could replace exact diagonalizations for fragment sizes
where it is not feasible while still providing highly correlated (fragment)
wave functions. Obviously, a careful analysis of the corresponding
symmetries should be carried out in each case. Work along these avenues is
in progress and will be reported elsewhere.

\label{conclusions}

\begin{acknowledgments}
This work was supported by the Department of Energy,
Office of Basic Energy Sciences, Grant No. DE-FG02-09ER16053. GES is a Welch Foundation
Chair (C-0036). Some of the calculations in
this work have been performed at the Titan computational facility, Oak Ridge
National Laboratory National Center for Computational Sciences, under
project CHM048. One of us (R.R-G.) would like to thank Prof. K. W. Schmid,
Institut f\"{u}r Theoretische Physik der Universit\"{a}t T\"{u}bingen, for
valuable discussions.
\end{acknowledgments}

\end{document}